\documentclass[a4paper,fleqn]{cas-dc}
    
\usepackage{amsmath,amssymb,amsfonts}
\usepackage{algorithmic}
\usepackage{graphicx}
\usepackage{textcomp}
\usepackage{xcolor}
\usepackage{xspace}
\usepackage{url}
\usepackage{colortbl}
\usepackage{booktabs}
\usepackage{paralist}
\usepackage{algorithm}
\usepackage{balance}
\usepackage{cleveref}
\usepackage{framed}
\usepackage{mdframed}
\usepackage{tabularx}
\usepackage{makecell}
\usepackage{threeparttable}
\usepackage{multirow} 
\usepackage[numbers]{natbib}
\usepackage{diagbox}
\usepackage{makecell,multirow,diagbox,rotating}
\usepackage{etoolbox} 
\usepackage{fancybox}
\usepackage{comment}
\usepackage[most]{tcolorbox}

\colorlet{gray1}{gray!70}
\colorlet{gray2}{gray!25}
\colorlet{gray3}{gray!15}
\definecolor{darkgrey}{HTML}{434343}
\definecolor{lightgrey}{HTML}{A9A9A9}
\definecolor{silver}{HTML}{D3D3D3}
\definecolor{midgrey}{HTML}{808080}
\definecolor{white}{HTML}{FFFFFF}
\definecolor{myblue1}{HTML}{0A5AA3}
\definecolor{myblue2}{HTML}{0F6FC6}
\definecolor{myblue3}{HTML}{A0B3DC}
\usepackage{caption}
\newtcolorbox{LYbox}[2][]{text width=0.95\linewidth,fontupper=\normalsize,
fonttitle=\bfseries\sffamily, colbacktitle=darkgrey,enhanced,
attach boxed title to top left={yshift=-2mm,xshift=3mm},
boxed title style={sharp corners},top=4pt,bottom=2pt,left=2pt,right=2pt,
title=#2,colback=white,coltitle=white}

\begin{document}
\makeatletter
\def\verbatim{\small\@verbatim \frenchspacing\@vobeyspaces \@xverbatim}
\makeatother
\renewcommand\arraystretch{1.4}
\let\WriteBookmarks\relax
\def\floatpagepagefraction{1}
\def\textpagefraction{.001}
\shorttitle{Enhancing the Accuracy and Comprehensibility in Architectural Tactic Detection}
\shortauthors{Cao et al.}

\title[mode = title]{Enhancing the Accuracy and Comprehensibility in Architectural Tactics Detection via Small Model-Augmented Prompt Engineering}



\author[a]{Lingli Cao}
\ead{cll@smail.nju.edu.cn}

\author[a]{He Zhang}
\ead{hezhang@nju.edu.cn}

\author[a]{Shanshan Li}
\ead{lss@nju.edu.cn}
\cormark[1]

\author[a]{Danyang Li}
\ead{522024320080@smail.nju.edu.cn}

\author[a]{Yanjing Yang}
\ead{yj_yang@smail.nju.edu.cn}

\author[a]{Chenxing Zhong}
\ead{chenxingzhong1125@gmail.com}

\author[a]{Xin Zhou}
\ead{xinzhou@nju.edu.cn}

\author[b]{Yue Xie}
\ead{yx388@cam.ac.uk}

\address[a]{State Key Laboratory of Novel Software Technology, Software Institute, Nanjing University, China}
\address[b]{Department of Engineering, University of Cambridge, UK}

\cortext[cor]{Corresponding author. Tel: +86-025-83621360-908.}

\begin{abstract}
Architectural tactics (ATs), as the concrete implementation of architectural decisions in code, address non-functional requirements of software systems. Due to the implicit nature of architectural knowledge in code implementation, developers may risk inadvertently altering or removing these tactics during code modifications or optimizations. Such unintended changes can trigger architectural erosion, gradually undermining the system's original design. While many researchers have proposed machine learning-based methods to improve the accuracy of detecting ATs in code, the black-box nature and the required architectural domain knowledge pose significant challenges for developers in verifying the results. Effective verification requires not only accurate detection results but also interpretable explanations that enhance their comprehensibility. However, this is a critical gap in current research. Large language models (LLMs) can generate easily interpretable ATs detection comments if they have domain knowledge. Fine-tuning LLMs to acquire domain knowledge faces challenges such as catastrophic forgetting and hardware constraints. Thus, we propose Prmt4TD, a small model-augmented prompting framework to enhance the accuracy and comprehensibility of ATs detection. Combining fine-tuned small models with In-Context Learning can also reduce fine-tuning costs while equipping the LLM with additional domain knowledge. Prmt4TD can leverage the remarkable processing and reasoning capabilities of LLMs to generate easily interpretable ATs detection results. Our evaluation results demonstrate that Prmt4TD achieves accuracy (\emph{F1-score}) improvement of 13\%-23\% on the ATs balanced dataset and enhances the comprehensibility of the detection results.


\end{abstract}


\maketitle

\section{Introduction}
\label{sec: introduction}

Software architecture is a set of architectural design decisions (ADDs) \cite{10.1007/978-3-540-24769-2_14}. ADDs (partially) fulfill one or more requirements of the given architecture. When addressing the quality requirements of software systems, architectural tactics (ATs) serve as the concrete implementation of architectural decisions. For example, build separation tactic can be employed to address scalability requirements in microservice applications \cite{8705256}. However, as systems evolve, architectural erosion has become a common occurrence \cite{296790}\cite{6365644}\cite{7270338}. One contributing factor to this is the inconsistency between the ADDs and the source code \cite{10.1145/1985793.1985942}\cite{9462989}. For example, to optimize performance, the architectural design requires the use of the pooling tactic to reduce the overhead caused by the frequent creation and destruction of objects. However, the code implementation directly creates new resource instances each time without reusing existing resources. This mismatch between design and implementation may lead to excessive resource consumption, increased system performance bottlenecks, and even resource exhaustion. Typically, the consistency between ADDs and code implementation relies on developers manually tracing the code to ADDs, which requires significant time and effort \cite{9825785}. Moreover, developers' level of domain knowledge about architecture greatly influences the quality and efficiency of tracing.

Several tools for managing architectural knowledge have been proposed \cite{RAZAVIAN2023111560}\cite{10.1145/3344948.3344960}\cite{4133434}. However, these tools only provide a limited record of abstracted architectural decisions at the design level and do not establish traceability links to specific code implementations, thereby offering limited theoretical support to developers during the system evolution phase. Subsequently, researchers have attempted to leverage machine learning (ML) techniques to automate the detection of ATs in code, thereby reducing labor costs while enabling traceability from the code to the underlying design decisions \cite{7270338}\cite{10.1007/978-3-030-58923-3_15}. However, the implementation of ATs in code does not follow a unified standard format \cite{9825785}\cite{Sharifi2023}. The existing ML-based approaches for AT detection typically only provide predicted labels along with prediction probabilities without any additional contextual information. As a result, while these methods can provide preliminary ATs detection results, manual verification is still required to ensure their correctness \cite{7930206}\cite{7270338}. For example, Mirakhorli et al. proposed a solution by detecting ATs in code, tracing them back to requirements, and visualizing them to aid developers in understanding underlying ADDs knowledge \cite{7270338}. However, the generated detection results require validation by architects or developers one by one. This introduces extra workload and necessitates that developers possess knowledge of ADDs to effectively validate the results.

The black-box detection largely overlooks the importance of comprehensibility for manual verification. For example, when a detection method labels a piece of code as implementing an "audit" tactic, developers may question the accuracy of the results. However, by providing clear reasoning such as "The code contains logic for recording user actions, detecting keywords like 'logEntry,' 'auditTrail,'" it not only helps developers quickly understand the design intent and objectives but also increases their confidence in the results, enabling them to make informed decisions based on this information. Automated AT detection is not intended to replace manual inspection, as even the most advanced artificial intelligence technologies cannot achieve 100\% accuracy in detection. To achieve effective human-machine collaboration, the results of automated ATs detection should be clear and well-reasoned to facilitate developer understanding. This means that good comprehensibility requires automated ATs detection to accurately identify ATs in the code and provide reasonable explanations for the judgments made. Recently, large language models (LLMs) have demonstrated the potential to address complex problems and enhance interpretability in various application scenarios due to their excellent reasoning and generative capabilities \cite{10.1145/3695988}\cite{10.1145/3695993}. For instance, Yang et al. \cite{DBLP:journals/jss/YangZMXYZSZ25} proposed using enhanced context-prompting techniques to detect vulnerabilities and provide explanations for detection results. Despite the large-scale pretraining data that LLMs have learned from, to achieve positive performance in specific downstream tasks, large-scale supervised data must be used for effective fine-tuning of LLMs or customized prompt engineering techniques \cite{zhao2024surveylargelanguagemodels}. However, the full parameter fine-tuning method is limited by data availability and resource constraints, rendering it inefficient and unsustainable \cite{han2024parameterefficientfinetuninglargemodels}. Furthermore, it may compromise the generalization ability of LLMs \cite{doi:10.1073/pnas.1611835114}. While more generic parameter-efficient fine-tuning (PEFT) techniques can enhance performance on downstream tasks, they often lead to catastrophic forgetting \cite{luo2025empiricalstudycatastrophicforgetting}\cite{doi:10.1073/pnas.1611835114}, thereby affecting the comprehensibility of the results. Therefore, it is necessary to implement customized prompt engineering in specific downstream tasks to achieve the accuracy and comprehensibility of the results.

However, prompting is limited by the context length, often providing insufficient domain knowledge \cite{xu2023small}. Thus, we propose a small model-augmented prompting framework for ATs detection, Prmt4TD, which employs two prompt engineering techniques: in-context learning (ICL) \cite{NEURIPS2020_1457c0d6} and chain-of-thought (CoT) \cite{DBLP:conf/nips/Wei0SBIXCLZ22}. The core idea of Prmt4TD is to utilize the implementation code of the target ATs to train small models to provide additional domain knowledge in ICL and reduce fine-tuning costs. CoT prompts guide LLMs in generating an interpretable process tailored to the current task, improving the transparency of automated ATs detection. Specifically, Prmt4TD uses Conformal Prediction (CP) \cite{vovk2005algorithmic}\cite{li-etal-2024-traq} to obtain a set of predicted labels containing the true label of the input code, with 95\% confidence. Then, based on lexical, syntactic, and semantic similarity, Prmt4TD samples candidate code snippets that are most similar to the input code from the predicted label set. The trained small model is then used to obtain the predicted labels and confidence scores of the candidate code snippets and the input code. The candidates, along with their predicted labels, confidence scores, and true labels, as well as the input code and its predicted labels and confidence levels, together form the ICL prompt. On the other hand, Prmt4TD uses the predicted label set generated by CP to fill in a predefined CoT template. We validate the effectiveness of Prmt4TD across various datasets. In terms of accuracy, Prmt4TD outperforms baseline methods by 13\%-23\% in \emph{F1-score} on the ATs balanced dataset. In the Hadoop case, Prmt4TD achieves an improvement of 4\%-23\% in \emph{F1-score} over baseline methods. Regarding comprehensibility, Prmt4TD achieves the highest clarity, surpassing baseline methods by the highest \emph{Clear} rate and lowest \emph{Unclear} rate. Additionally, we conduct an ablation study, which yields positive feedback on Prmt4TD.

The main contributions of this paper are as follows.

\begin{itemize}
    \item This paper proposes a generalizable LLMs prompting framework, Prmt4TD, for automated ATs detection. This method combines the advantages of small models and LLMs with the potential to generalize to various ADDs detection tasks.
    \item We demonstrate the effectiveness of using Prmt4TD for automating ATs detection in the dataset, compared to current state-of-the-art methods.
    \item Through practical large-scale projects, we evaluate the advantages of Prmt4TD in automating ATs detection tasks, providing comprehensibility for black-box detection, and enhancing the transparency of ADDs.  
    
\end{itemize}

The rest of the paper is organized as follows. \Cref{sec:related} reviews the motivation and related work. \Cref{sec: framework} describes the proposed prompting framework Prmt4TD. \Cref{sec:experiments} and \Cref{sec:analyis} present the experimental design and results respectively. \Cref{sec:discussion} discusses the findings. \Cref{sec:ttv} lists the threats to the validity of the study. Finally, \Cref{sec:conclusion} concludes the paper and points out directions for future research.

\section{Motivation and Related Work}
\label{sec:related}
\subsection{Motivation}
ATs are concrete implementations of design decisions to achieve specific quality attributes in a system. For example, security tactics and patterns are used to mitigate security threats \cite{10.1145/2993412.3007552}. However, without a clear understanding of the underlying design decisions, developers are likely to incorrectly implement or modify code related to ATs. This misalignment between ATs and the code can contribute to the deterioration of the architecture over time. Existing ML-based ATs detection methods primarily focus on the accuracy of detection results. These methods are black-box approaches that lack transparency in the detection process, making it difficult for developers to trust the results when only the predicted label is provided. The detection results still require significant time and ADD knowledge for manual verification. 

The development of LLMs offers hope in addressing these challenges. LLMs have demonstrated excellent performance in various software engineering tasks such as automated code review \cite{10.1145/3695993}, vulnerability detection \cite{DBLP:journals/jss/YangZMXYZSZ25}\cite{lu2024grace}, and code comment generation \cite{geng2024large}, due to their excellent reasoning and generative capabilities \cite{10.1145/3695988}. LLMs can generate easily interpretable ATs detection comments if they have domain knowledge. The main approaches for applying LLMs to downstream tasks include fine-tuning and prompt engineering. Full parameter fine-tuning techniques are constrained by data and computational resources, are relatively inefficient, and may harm the model's generalization ability \cite{han2024parameterefficientfinetuninglargemodels}\cite{doi:10.1073/pnas.1611835114}. In contrast, commonly used PEFT methods, while alleviating some of these limitations, often lead to catastrophic forgetting \cite{luo2025empiricalstudycatastrophicforgetting}\cite{doi:10.1073/pnas.1611835114}. Therefore, it is necessary to implement customized prompt engineering in specific ATs detection tasks to achieve the accuracy and comprehensibility of the results.

However, prompting is limited by the context length, often providing insufficient domain knowledge \cite{xu2023small}. Therefore, we propose combining the superICL technique with COT prompts to effectively leverage the reasoning and generation capabilities of LLMs for ATs detection tasks. Conventional ICL techniques typically provide limited supervised demonstrations to guide the LLM in learning downstream tasks. In contrast, superICL enables fine-tuned smaller models to offer additional task-specific information to the LLM and enhances the interpretability of these smaller models \cite{xu2023small}. The integration of these smaller models with the LLM through ICL has been shown to achieve superior performance in supervised tasks \cite{xu2023small}\cite{DBLP:journals/jss/YangZMXYZSZ25}. The CoT technique can effectively guide the reasoning abilities of LLMs, helping the model to better understand tasks and generate supplementary natural language text that assists in task resolution \cite{chan2023knife} \cite{zhao2024surveylargelanguagemodels}. Additionally, the selection of appropriate demonstrations and context length in ICL can significantly influence the performance of prompting \cite{randl2024cicle}\cite{10.1162/tacl_a_00638}. To address this, we propose utilizing CP, a statistical technique that associates the predictions of a classification algorithm with confidence levels, aiming to construct a minimized prediction set \cite{vovk2005algorithmic}\cite{li-etal-2024-traq}. When applied to multi-task classification, this approach helps to maintain the context as concisely and resource-efficient as possible \cite{randl2024cicle}.

\subsection{Related Work}
Research has demonstrated that even with a perfectly designed architecture that meets the current system's quality requirements, improper implementation or maintenance of architectural tactics by developers can lead to architectural erosion \cite{7180098}\cite{4343775}. Although many architectural knowledge management tools have been proposed, they often fail to establish traceability links between architectural decisions and concrete implementation code. As a result, developers must still explore low-level code implementations based on high-level architectural knowledge \cite{10.1145/1985793.1985942}. The most relevant work in this regard is the method proposed by Mirakhorli and Cleland-Huang \cite{7270338}. They use ML to automatically detect ATs in code by training a custom classifier on architecturally significant code snippets extracted from open-source systems, identifying 10 common tactics. Subsequently, Mujhid et al. proposed a search engine designed to find and reuse code related to ATs, thereby supporting development efforts \cite{MUJHID201781}. Sharifi and Abdolahzadeh-Barforoosh introduced a novel approach to recognizing architectural tactics using clone microtactics \cite{Sharifi2023}. With language model excelling at processing text tasks \cite{ALHOSHAN2023107202}\cite{10.1145/3551349.3560417}\cite{9218141}, Okutan et al. \cite{okutan2023novelapproachidentifysecurity} extracted tactic and non-tactic code snippets from Stack Overflow. By combining these snippets with the BERT model, they trained a classification model capable of identifying security tactics code with high confidence. Keim et al. attempted to detect ATs in code by fine-tuning BERT. Although their results are not entirely satisfactory, the study demonstrates the significant potential of applying language models to such problems \cite{10.1007/978-3-030-58923-3_15}.

However, existing automated ATs detection works primarily focus on improving the accuracy of detecting ATs in code while often neglecting the comprehensibility of the results. As a result, developers still need to spend considerable time and effort verifying the validity of the detection outcomes. This paper proposes Prmt4TD, a method that combines the strengths of small models and LLMs, aiming to enhance the comprehensibility of detection results while ensuring accuracy.

\begin{figure*}[htbp]
    \begin{center}
    \includegraphics[width=1.0\textwidth]{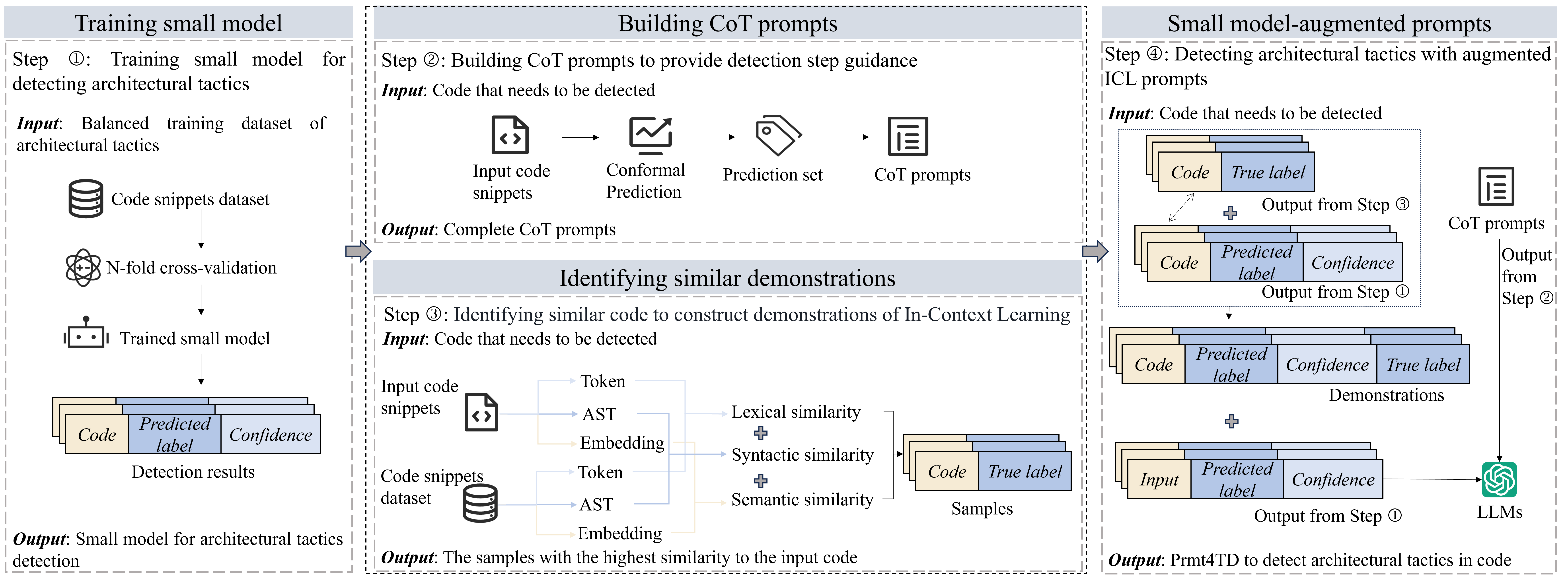}
    \vspace{-2.0ex}
    \end{center}
    \caption{An overview of the proposed Prmt4TD}
    \vspace{-4.0ex}
\label{fig: Overview}
\end{figure*}

\section{The Proposed Framework: Prmt4TD}\label{sec: framework}
The proposed Prmt4TD framework, as shown in \Cref{fig: Overview}, consists of four stages: training the small model, building CoT prompts, identifying similar demonstrations, and small model-augmented prompting. The trained small model supports both the building of COT and the identification of similar demonstrations. Afterward, the results from these three stages are integrated to generate small model-augmented prompts. In this section, we will detail the specific workflow of each stage.

\subsection{Training The Small Model}

The core concept of Prmt4TD is to leverage ICL \cite{NEURIPS2020_1457c0d6} and CoT \cite{DBLP:conf/nips/Wei0SBIXCLZ22} techniques to design high-quality prompts, enabling LLMs to achieve satisfactory performance in the ATs detection task. To this end, Prmt4TD employs a small model as a front-end plugin for LLMs, enhancing the construction of ICL prompts and effectively stimulating the implicit fine-tuning capabilities of LLMs. This integration mitigates the risk of performance degradation in detection tasks caused by hallucination issues and data distribution discrepancies \cite{DBLP:journals/jss/YangZMXYZSZ25}. Consequently, the first step of Prmt4TD is to train a small model specifically tailored for ATs detection. Given a dataset of code snippets related to ATs, represented as 
\(\mathcal{D} = \{(x_i, y_i)\}_{i=1}^N,\) where \(y \in \{0, 1, 2, \ldots, n\}\), and \(n\) is the number of true labels minus one, \(x_i\) represents the source code of a class, and \(y_i\) is the ground truth (each number corresponds to a specific architectural tactic label). On this dataset, we can train a small model \( f_{\text{ML}}\) for ATs detection, which is defined as: 

\begin{equation}
\hspace{1.2cm}
 \begin{aligned}
    {f_{\text{SM}}(x_i) = (\hat{y}_i, c_i)}
    \end{aligned}
\label{eq1}
\end{equation}

where \(x_i \in X\) is the input feature representation of the code snippet, \(\hat{y}_i \in Y = \{0, 1, 2, \ldots, n\}\) is the predicted AT label, and \(c_i \in [0, 1]\) is the confidence of the model prediction for label \(\hat{y}_i\). The model \(f_{\text{ML}}(x_i)\) aims to minimize the loss function \(L = \frac{1}{N} \sum_{i=1}^N \ell(\hat{y}_i, y_i),
\) where \(\ell(\hat{y}_i, y_i)\) represents the loss between the predicted label \(\hat{y}_i\) and the ground truth \(y_i\).

\subsection{Buidling CoT Prompts}
\label{sec: framework-ICL}
\begin{figure}[htbp]
\begin{center}
    \includegraphics[width=1.0\linewidth]{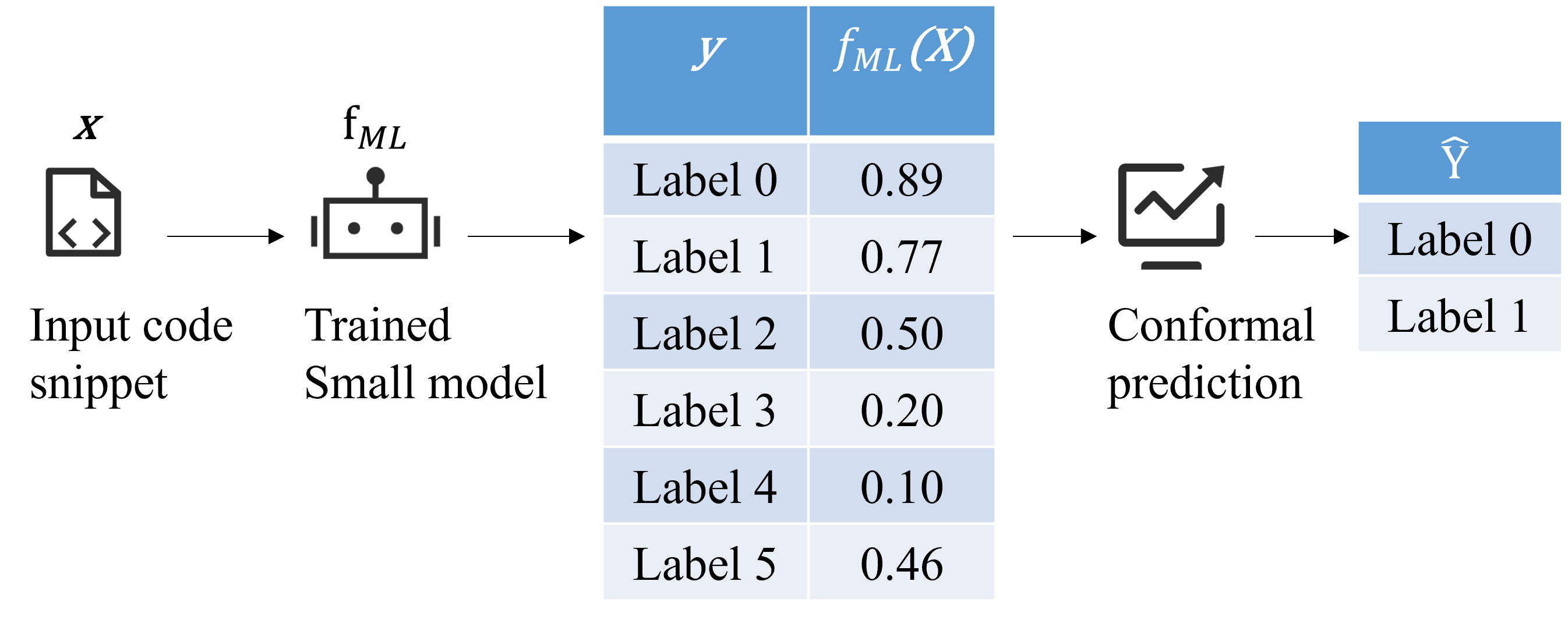}
    \vspace{-2.0ex}
\end{center}
\vspace{-4.0ex}
\caption{An example of CP process of Prmt4TD}
\vspace{-4.0ex}
\label{fig: CPexample}
\end{figure}

The second step of Prmt4TD is to generate a specified reasoning path for each input test code snippet using CoT. CoT is a prompting strategy. Unlike traditional prompts that use input-output pairs, CoT incorporates a series of intermediate reasoning steps into the prompt \cite{DBLP:conf/nips/Wei0SBIXCLZ22}. By emulating the human problem-solving approach, CoT decomposes complex problems into a series of smaller, simpler ones. This allows LLMs to focus on solving these simpler problems \cite{wang2023selfconsistencyimproveschainthought}. The construction of CoT prompts can be divided into two stages. First, the trained small model is applied to analyze and detect the test samples. Then, the Conformal Prediction (CP) \cite{li-etal-2024-traq}\cite{vovk2005algorithmic} is used to generate a set of predicted labels with user-defined confidence levels. The core idea of CP is to quantify the uncertainty of the classification model's predictions, providing a confidence guarantee for each prediction. Specifically, CP defines a non-conformity function to measure the discrepancy between the classification model's predicted output and the actual observation. The non-conformity function is then calibrated using a calibration set. After calibration, the model makes predictions on the input data, calculating the non-conformity scores for all candidate labels, thus generating a prediction set. This set ensures that the true label is included with the confidence level defined by the user. In the context of classification tasks, the stronger the classification model, the better the performance of its non-conformity function, and the smaller the prediction set generated by CP while maintaining the desired confidence level \cite{randl2024cicle}. As shown in \Cref{fig: CPexample}. This set of predicted labels provides potential answers, including confidence information, to LLMs, represented as: 

\begin{equation}
 \begin{aligned}
    {\hat{Y}_i = \{ y^{(j)} \mid \text{incof}_i^{(j)} \leq \text{quantile}_{1-\alpha} \left( \{\text{incof}_i^{(k)}\}_{k=1}^m \right) \}}
    \end{aligned}
\label{eq2}
\end{equation}

where \( \hat{Y}_i \) represents the set of all labels for the test sample \( x_i \) that satisfy the CP conditions, which meet the specified confidence requirement. \( y^{(j)} \) denotes a potential predicted label. \( \text{incof}_i^{(j)} \) represents the "non-conformity score" calculated by the CP method, which measures the inconsistency between the predicted label \( y^{(j)} \) and the test sample \( x_i \). And \( \text{quantile}_{1-\alpha} \left( \{\text{incof}_i^{(k)}\}_{k=1}^m \right) \) is the \( (1-\alpha) \)-quantile of the "non-conformity score" distribution, which indicates that we select those labels whose "non-conformity score" is less than or equal to this quantile to be part of the prediction set. For example, in this paper, we define \( \alpha = 0.05 \), so \( 1-\alpha = 0.95 \) means we choose the top 95\% of labels with lower "non-conformity score" as part of the final prediction set.

Next, we develop a generalized CoT template for detecting ATs detection tasks. As research on the application of LLMs to ATs detection tasks is still in its early stages, with limited references available, we construct this template by drawing on related CoT research \cite{liu2023not}\cite{DBLP:journals/jss/YangZMXYZSZ25} and the typical steps involved in manually and ML-based ATs detection. The template is structured as follows:

\begin{itemize}
    \item \textbf{Semantic understanding:} Understanding the core behavior and intent of the input code snippet.
    
    \item \textbf{Structural analysis:} Analyzing key terms and structures in the code to identify components that may implement a specific AT.

    \item \textbf{Classification judgment:} Based on the previous analysis and context, assessing whether the code snippet is related to a specific AT label.
    
    \item \textbf{Label identification:} If related, further identify the label of the code snippet from the prediction set. If not related, check if the label belongs to a category outside of the prediction set.
    
    \item \textbf{CoT Construction:} Integrating the reasoning steps and sequentially deriving whether the current detection sample belongs to the ATs detection task and determining its corresponding AT label.
\end{itemize}

\begin{figure}[htbp]
\begin{center}
    \includegraphics[width=1.0\linewidth]{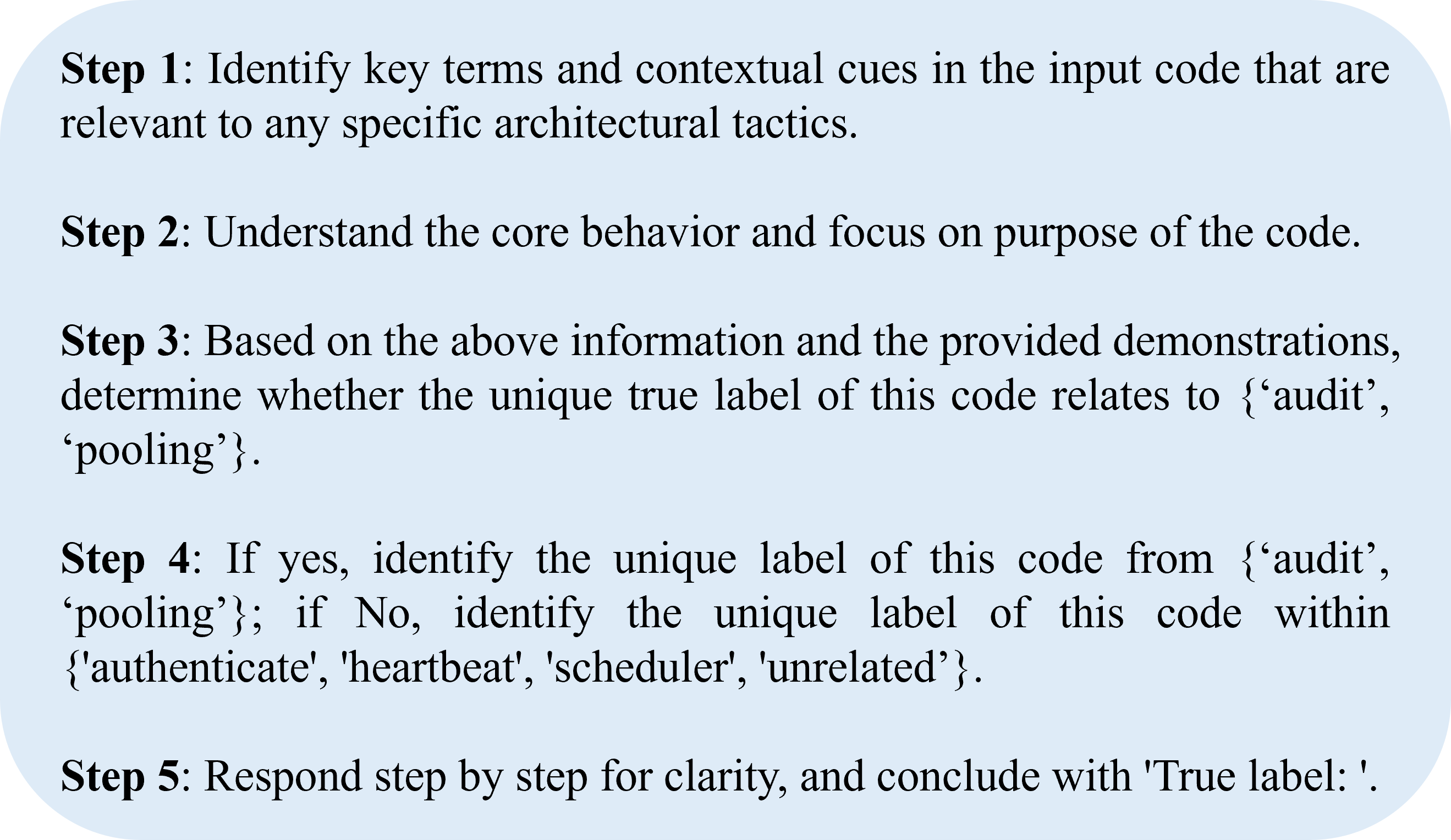}
    \vspace{-2.0ex}
\end{center}
\vspace{-4.0ex}
    \caption{An example of the specific COT prompts of Prmt4TD}
    \vspace{-2.0ex}
\label{fig: COTexample}
\end{figure}

Prmt4TD combines the results of CP with the CoT template to generate a custom CoT prompt for the current detection sample. For example, given a code snippet to be detected, if the detection result of the small model is the AT "audit" and the CP method provides a minimized set of predicted labels containing the true label with 95\% confidence, such as \{‘audit’, ‘pooling’\}, Prmt4TD generates a specific CoT prompt for this detection sample by combining the result of the small model with the CoT template, as shown in \Cref{fig: COTexample}.

\subsection{Identifying Similar Demonstrations}
\label{sec: framework-COT}

ICL, as a typical method for utilizing LLMs, how to design demonstrations effectively is an important research problem \cite{zhao2024surveylargelanguagemodels}. By leveraging the demonstrations provided in ICL, LLMs can generate meta-gradients related to the demonstrations and implicitly perform gradient descent through the attention mechanism, thereby achieving implicit fine-tuning of the LLMs \cite{zhao2024surveylargelanguagemodels}\cite{dai2022can}\cite{pmlr-v202-von-oswald23a}. Given a set of demonstrations, represented by \( \text{Demoi} = \{ f(x_1, y_1), \dots, f(x_i, y_i) \} \), where \( f(x_i, y_i) \) is the function that maps the \( i \)-th demonstration into a natural language prompt. Using ICL to guide LLMs in generating task-specific predictions can be represented as:  

\begin{equation}
 \begin{aligned}
    {F_{\text{LLM}}(I, \text{Demo\_i}, f(x_{i+1}, \_)) = \hat{y}_{i+1}}
    \end{aligned}
\label{eq3}
\end{equation}

where \( I \) is the description of the task, \( x_{i+1} \) is the input of the question to the LLM, and \(\hat{y}_{i+1}\) is the prediction generated by the LLMs based on ICL.

Many studies have shown that the performance of ICL heavily depends on the quality of the demonstrations, including the selection of examples, the formatting of the examples, and the ordering of the demonstrations \cite{zhao2024surveylargelanguagemodels}\cite{dong-etal-2024-survey}. To provide high-quality demonstrations for ATs detection tasks, we propose a heuristic method to retrieve examples relevant to \( x_{i+1} \). Specifically, we calculate similarity scores based on semantic, lexical, and syntactic similarities between \( x_{i+1} \) and the examples in the sample set. These scores are then integrated to identify the example code most similar to \( x_{i+1} \). Previous research on code similarity mainly focuses on lexical and syntactic aspects, often overlooking semantic differences, which may lead to the selection of code snippets that are syntactically similar but semantically distinct \cite{lu2024grace}. Therefore, we adopt a comprehensive approach that considers lexical, syntactic, and semantic similarities when analyzing code snippet similarity in order to address this challenge.

We extract the semantic embedding vector, token set, and abstract syntax tree (AST) of the code to capture its functional, lexical, and structural features, respectively. Based on these features, we calculate the semantic, lexical, and syntactic similarities between the current code and the code snippets in the example set. Finally, we select the code snippet most similar to the current code from the example set. Specifically, we measure semantic similarity by computing the cosine similarity between the semantic embedding vectors of the current code and the example code. Cosine similarity is a widely used metric in various domains, including text and image analysis \cite{pedregosa2011scikit}. The computation is expressed in \Cref{eq4}, where \(\mathcal{V}_c\) and \(\mathcal{V}_i\) represent current code and example code semantic embedding vectors, respectively. \(\mathcal{V}_c \cdot \mathcal{V}_i\) is the dot product of the two vectors, and \(\|\mathcal{V}_c\|\) and \(\|\mathcal{V}_i\|\) are their respective \(L_2\)-norms:  
\begin{equation}
 \begin{aligned}
    {\text{Semantic\_similarity}(\mathcal{V}_c, \mathcal{V}_i) = \frac{\mathcal{V}_c \cdot \mathcal{V}_i}{\|\mathcal{V}_c\| \|\mathcal{V}_i\|}}
    \end{aligned}
\label{eq4}
\end{equation}

For lexical similarity, we calculate the Jaccard similarity \cite{jain1988algorithms} between the lexical token sets of the current code and the example code. The jaccard similarity is commonly used to assess set similarity by evaluating the proportion of the intersection to the union of two sets \cite{Ragkhitwetsagul2018}. \Cref{eq5} reflects the degree of lexical overlap between code snippets, where \(\mathcal{T}_c\) and \(\mathcal{T}_i\) represent current code and example code token sets, respectively.

\begin{equation}
 \begin{aligned}
    {\text{Lexical\_similarity}(\mathcal{T}_c, \mathcal{T}_i) = \frac{|\mathcal{T}_c \cap \mathcal{T}_i)|}{|\mathcal{T}_c \cup \mathcal{T}_i)|}}
    \end{aligned}
\label{eq5}
\end{equation}

In terms of syntactic similarity \cite{Ragkhitwetsagul2018}, we first parse the code to generate the corresponding AST. Then, the AST is converted into a node sequence, and the Levenshtein distance between these sequences is calculated to quantify syntactic similarity. The Levenshtein distance is widely adopted in research to handle AST relationships. The formula is presented below, where \(\mathcal{S}_c\) and \(\mathcal{S}_i\) represent current code and example code sequences, respectively.

\begin{equation}
 \begin{aligned}
    {\text{Syntactic\_similarity}(\mathcal{S}_c, \mathcal{S}_i) = 1 - \frac{\text{Lev}(\mathcal{S}_c, \mathcal{S}_i)}{\max(\text{len}(\mathcal{S}_c), \text{len}(\mathcal{S}_i))}
}
    \end{aligned}
\label{eq6}
\end{equation}

To comprehensively measure the similarity between code snippets, we integrate three types of similarity, as defined in \Cref{eq7}, where \(\alpha + \beta + \gamma = 1\).

\begin{equation}
    \begin{aligned}
        \text{similarity\_score} &= \alpha \cdot \text{Semantic\_similarity} \\
        &+ \beta \cdot \text{Lexical\_similarity} \\
        &+ \gamma \cdot \text{Syntactic\_similarity}
    \end{aligned}
\label{eq7}
\end{equation}

 In Prmt4TD, the weights represented by the three coefficients are assigned equal importance. Based on the similarity score, we select the code snippet most similar to the current code from the example set to serve as a demonstration in the ICL prompt. It is worth noting that during the selection process, we prioritize choosing the most similar code snippet from the example set corresponding to the predicted set provided by CP.

\begin{figure}[htbp]
\begin{center}
    \includegraphics[width=1.0\linewidth]{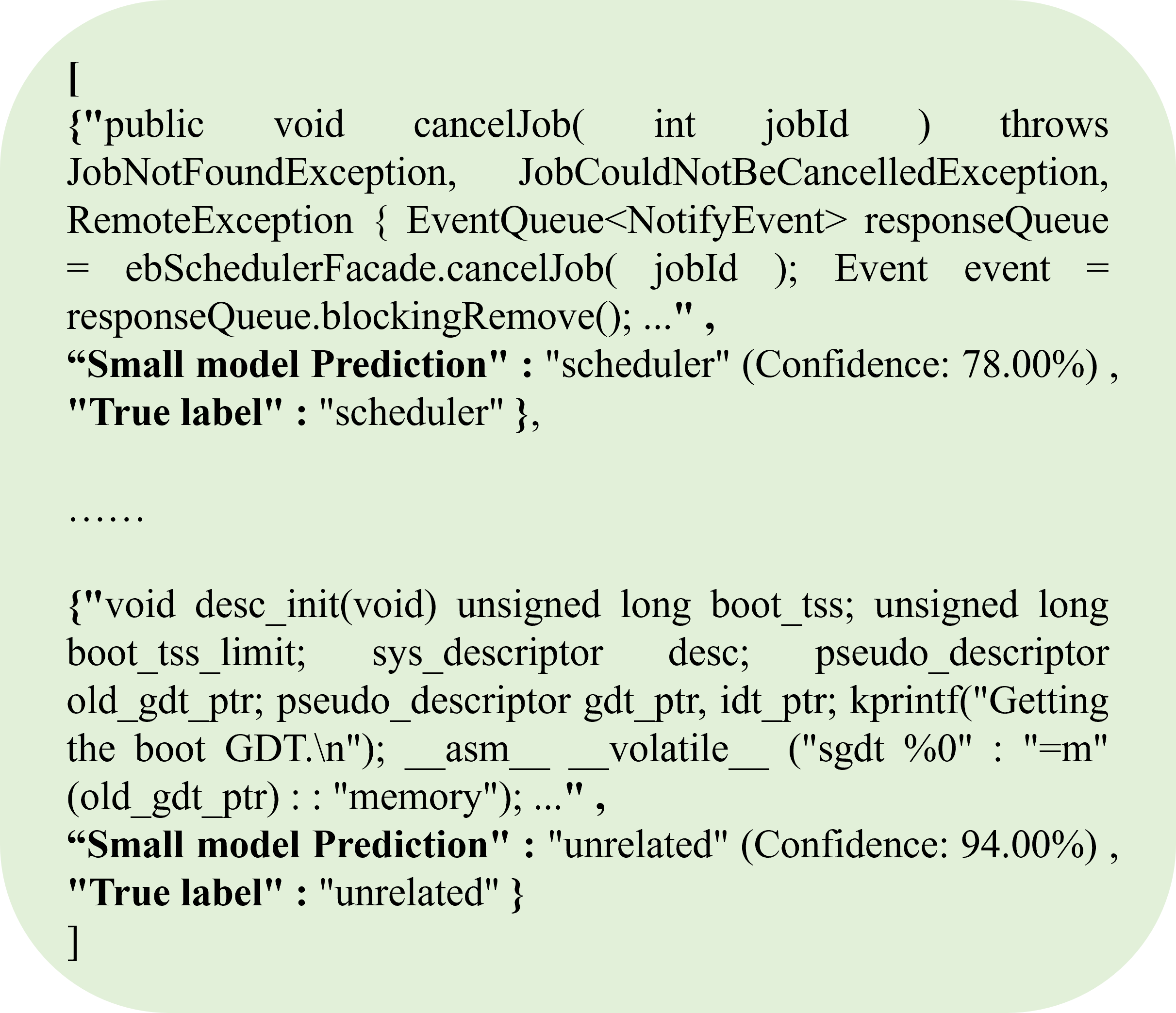}
    \vspace{-2.0ex}
\end{center}
\vspace{-4.0ex}
\caption{An example of demonstrations in ICL prompt of Prmt4TD}
\vspace{-2.0ex}
\label{fig: demonstration}
\end{figure}

 After selecting the demonstrations, we need to format them to integrate into the ICL prompt. First, we detect code snippets using the ML model trained in the first step, identifying their predicted labels and confidence scores. Then, we combine the code snippets, the results detected by the small model, and the true labels to form the formatted demonstrations. Finally, we sort the demonstrations in the ICL according to the order of the labels in the predicted set provided by CP, from most similar to least similar (an example is shown in \Cref{fig: demonstration}).

\subsection{Small Model-augmented Prompting}
\label{sec: Prompts Synergy}
\begin{figure}[htbp]
\begin{center}
    \includegraphics[width=1.0\linewidth]{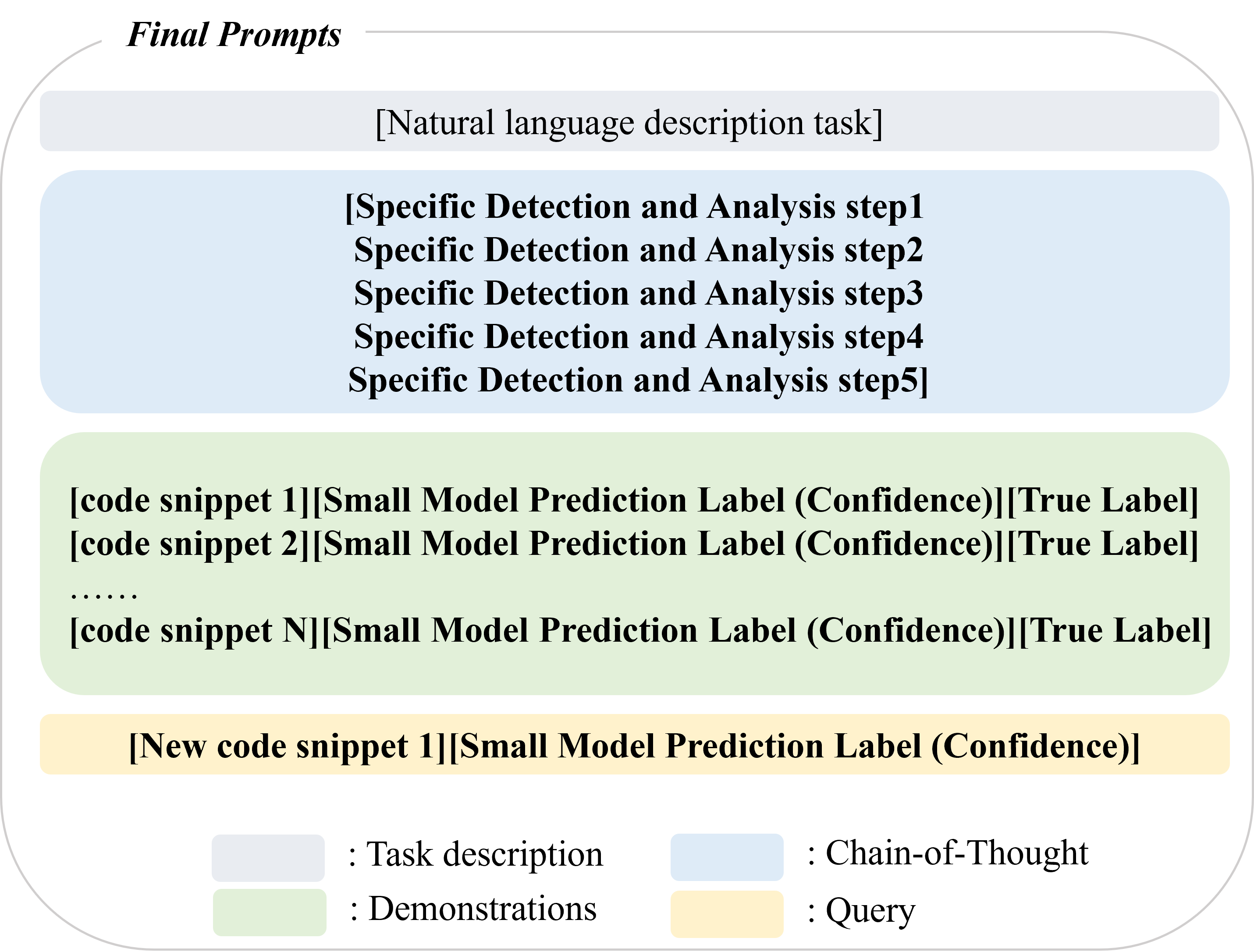}
    \vspace{-2.0ex}
\end{center}
\vspace{-4.0ex}
    \caption{The final prompts template of Prmt4TD}
    \vspace{-2.0ex}
\label{fig: finalprompt}
\end{figure}

After completing the customization of CoT prompts and the selection of demonstrations for ICL, we can proceed to integrate these components into a final prompt design tailored for ATs detection. As shown in \Cref{fig: finalprompt}, the proposed prompting framework, Prmt4TD, consists of two main components: the basic task description prompt and the enhanced ICL prompt. When deploying LLMs for downstream tasks, it is generally necessary to provide a basic prompt to clarify the task objectives. However, a basic prompt alone is insufficient \cite{achiam2023gpt}. To improve the model's task comprehension, it is essential to provide task-relevant contextual information that equips LLMs with the background knowledge needed to find patterns and learn to make effective predictions \cite{xu2023small}. 

Building on this foundation, we further incorporate CoT prompts and the prediction results with confidence scores from small models to enable LLMs to make final predictions along with reasoning explanations. The CoT prompts help LLMs establish a clear reasoning pathway, thereby enhancing the transparency of the prediction process. Meanwhile, the prediction labels provided by small models facilitate the LLMs' understanding of the relationship between demonstration samples, task-specific knowledge, and ground truth labels. During task execution, LLMs consider the small model's predicted labels and their confidence levels to dynamically decide whether to follow the predictions or rely on their reasoning. It improves the overall performance of LLMs in ATs detection tasks.

\section{Experimental Design}\label{sec:experiments}
This section provides the foundational information for the experiments, including the research question, datasets, baseline methods, and evaluation metrics.

\subsection{Research Questions}
We evaluate the performance of Prmt4TD in terms of accuracy and comprehensibility compared to advanced baseline methods through the following three research questions (RQs). 

\smallskip
\begin{itemize}
    \item [\textbf{RQ1:}] \textbf{Which category of small models is most effective for Prmt4TD?}
\end{itemize} 

\noindent\emph{\textbf{Motivation:}}
Prmt4TD integrates the prediction labels, confidence scores, and true label information provided by small models in the ATs detection task into the ICL process. This approach leverages the knowledge learned by small models in ATs detection to enhance the performance of LLMs on this task. In this RQ, we aim to identify which category of small model is suitable for such tasks. We compare three representative ML models: Support Vector Machine (SVM) \cite{joachims1999transductive}, Adaboost \cite{freund1996experiments}, and Bagging \cite{breiman1996bagging} and two deep learning (DL) models: BERT \cite{devlin2019bertpretrainingdeepbidirectional} and CodeBERT \cite{feng2020codebertpretrainedmodelprogramming} (cf. \Cref{ex: ML models} for rationales).

\smallskip
\begin{itemize}
    \item [\textbf{RQ2:}] \textbf{How accurately does Prmt4TD perform in detecting ATs tasks?} 
\end{itemize} 

\noindent\emph{\textbf{Motivation:}}
From a practical perspective, an effective automated ATs detection method should first ensure high accuracy in its detection results. On this basis, it is reasonable to discuss the comprehensibility of the results further. In this RQ, we aim to explore whether Prmt4TD can enhance the accuracy of ATs detection by leveraging the strengths of small models and LLMs. To this end, we compare Prmt4TD with two state-of-the-art baseline methods: Tactic Det. \cite{7270338} and BERT4TD\cite{10.1007/978-3-030-58923-3_15}, and three common ML models for detecting ATs: SVM \cite{joachims1999transductive}, Adaboost \cite{freund1996experiments}, and Bagging \cite{breiman1996bagging} to assess whether it outperforms these baselines.

\smallskip
\begin{itemize}
    \item [\textbf{RQ3:}] \textbf{How does Prmt4TD perform in providing comprehensible reasoning explanations for ATs detection tasks?} 
\end{itemize} 

\noindent\emph{\textbf{Motivation:}}
Previous research has primarily focused on using ML-based methods for detecting ATs, often overlooking the comprehensibility of the detection results. This limitation may undermine developers' trust in the detection outcomes. In this RQ, we aim to explore whether Prmt4TD can enhance the comprehensibility of detection results. We evaluate the comprehensibility of the generated detection comments through human assessments (cf. \Cref{ex: comprehensibility} for rationales). It is important to emphasize that the evaluation of comprehensibility is based on accurate ATs detection. Therefore, in RQ3, the accuracy of ATs detection using different prompting frameworks must also be assessed.

\smallskip
\begin{itemize}
    \item [\textbf{RQ4:}] \textbf{How do the different components contribute to the effectiveness of Prmt4TD?} 
\end{itemize} 

\noindent\emph{\textbf{Motivation:}}
In this RQ, we conduct ablation experiments to analyze the impact of each component in Prmt4TD. Specifically, we examine three key components within Prmt4TD:
\begin{itemize}
    \item \textbf{Small model:} Providing the prediction during demonstrations, as well as the prediction for the input code.
    \item \textbf{CoT:} Guiding the specific ATs detection reasoning process of LLMs.
    \item \textbf{Demonstrations:} Providing Similar demonstrations for LLMs to detect ATs in the input code.
\end{itemize}

\subsection{Datasets}
\label{ex: datasets}
Although ATs are a fundamental component of architectural knowledge, offering a systematic set for architectural design decisions, and are the primitive solutions of architectural patterns \cite{MARQUEZ2023111558}. Current research on ATs has, to some extent, overlooked the gap between tactics and code implementation. This gap may hinder the practical application of ATs. At present, studies in this field mainly rely on a balanced dataset compiled by  Mirakhorli et al. \cite{7270338}, which focuses on Java code. This dataset encompasses multiple ATs and, for each tactic, includes 50 code snippets related to the tactic and 50 unrelated ones extracted from 50 open-source projects implementing the tactic. Our research aims to detect five ATs included in the dataset: \emph{Audit Trail}, \emph{Authentication}, \emph{Heartbeat}, \emph{Resource Pooling}, and \emph{Scheduling} (as shown in \Cref{tab: tactics}). On this basis, we conduct a case study to evaluate the detection performance of Prmt4TD in the Hadoop Distributed File System (HDFS)\footnote{\url{https://hadoop.apache.org}}, with the case data sourced from \cite{7270338}\cite{10.1007/978-3-030-58923-3_15}. HDFS is a widely used distributed software system comprising over 1,700 classes, primarily providing functionalities for distributed storage and large-scale data processing. The ATs information identified in this case study is presented in \Cref{tab: datasethadoop}.

\begin{table}[htbp]
\scriptsize
\centering
\renewcommand\arraystretch{1.25}
\caption{\textbf{Description of the architectural tactics (cf. \cite{bass2012software})}}
\label{tab: tactics}
\begin{tabular}{p{1.5cm}p{6cm}}
\hline
\textbf{Tactic Name} & \textbf{Description} \\ \hline
Audit Trail & It records and tracks activities, maintaining a copy of each transaction and its associated identifying information. \\ 
Authentication & It verifies the authenticity of a user or system entity's identity. Authentication is typically achieved through methods such as passwords, digital certificates, or biometric recognition. \\
Heartbeat & It periodically sends signals to monitor the operational status of its components. If the heartbeat fails, it assumes the originating component has failed. \\
Pooling & It shares and reuses system resources. Pooling is commonly used to share threads, database connections, sockets, and other similar resources. \\
Scheduling & Resource contention is managed through the scheduling tactic, which optimizes task execution order and improves resource utilization.  \\ \hline
\end{tabular}
\vspace{-0.0ex}
\end{table}

\begin{table}[htbp]
\scriptsize
\centering
\renewcommand\arraystretch{1.25}
\caption{\textbf{ Information of architectural tactics in Apache Hadoop}}
\label{tab: datasethadoop}
\begin{tabular}{p{1.5cm}p{4.5cm}p{1cm}}
\hline
\textbf{Tactic Name}    & \textbf{Package Name} & \textbf{Classes} \\ \hline
Audit Trail      & mapred package                                                        & 4       \\
Authentication   & security package, MapReduce, and HDFS Subsystem                       & 25      \\
Heartbeat        & MapReduce Subsystem, HDFS Subsystem                                   & 24      \\
Pooling & mapred package, compress package, HDFS subsystem, MapReduce subsystem & 85      \\
Scheduling       & common, MapReduce                                                     & 68     
\\ \hline
\end{tabular}
\vspace{-0.0ex}
\end{table}

\subsection{Baselines}
\label{ex: baselines}
\subsubsection{Baselines for Evaluating Accuracy}
\label{ex: baselinesML}
We compare Prmt4TD with two advanced methods \cite{7270338}\cite{10.1007/978-3-030-58923-3_15} and three common ML models \cite{joachims1999transductive}\cite{freund1996experiments}\cite{breiman1996bagging} for ATs detection to assess its accuracy. These baseline methods have gained widespread recognition in the field of ATs detection. Furthermore, these methods utilize datasets that are identical or similar to the one employed by Prmt4TD during evaluation and are either publicly available or reproducible.

\begin{enumerate}[0]
\item[$\bullet$ \textbf{Tactic Det.}:] Tactic Det. \cite{7270338} primarily relies on information retrieval (IR) and ML techniques to train a custom classifier. This classifier is trained using code extracted from 50 open-source software systems that are performance-centric and safety-critical, with the key goal of identifying specific terms commonly found in the implementation of ATs.

\item[$\bullet$ \textbf{BERT4TD}:] BERT4DT \cite{10.1007/978-3-030-58923-3_15} treats the ATs detection as a text classification problem. It utilizes a ten-fold cross-validation approach to fine-tune the BERT model for detecting ATs in code. The results demonstrate the significant potential of applying language models in this domain.

\item[$\bullet$ \textbf{SVM}:] SVM \cite{joachims1999transductive} is a supervised learning model used for classification and regression analysis. SVM aims to find a hyperplane that maximizes the margin between different classes. In high-dimensional spaces, SVM can handle nonlinear classification problems using kernel functions (such as linear kernel, polynomial kernel, RBF kernel).

\item[$\bullet$ \textbf{Adaboost}:] Adaboost \cite{freund1996experiments} is a boosting technique that constructs a high-performing classifier by combining multiple weak classifiers using a voting mechanism. In classification tasks of higher complexity, AdaBoost has exhibited superior performance.

\item[$\bullet$ \textbf{Bagging}:] Bagging \cite{breiman1996bagging} is a parallel ensemble learning method. It generates multiple subsets from the original dataset by performing random sampling with replacement and trains different classifiers on each of these subsets. The final prediction is obtained by aggregating the predictions of the individual classifiers through a majority voting scheme. 
\end{enumerate}

\subsubsection{Baselines for Evaluating Comprehensibility}
\label{sec:baseline for Comprehensibility}
Our objective is to evaluate Prmt4TD's ability to address the issue of comprehensibility. As previously discussed, existing ATs detection methods have overlooked the comprehensibility of detection results. Therefore, in assessing the comprehensibility provided by Prmt4TD, we deliberately exclude the influence of these methods. Here, we validate comprehensibility by comparing the differences between the detection comments generated by Prmt4TD and those generated by the baseline prompting framework.

\begin{enumerate}[0]
\item[$\bullet$ \textbf{P}$_{\text{Bas}}$](Basic prompt): When using LLMs for ATs detection tasks, it is essential to provide a basic prompt that clearly describes the objective of the detection task and specifies the desired output format \cite{NEURIPS2020_1457c0d6}.

\begin{tcolorbox}[width=\linewidth,boxrule=1.0pt, left=3pt, right=3pt, top=3pt, bottom=3pt, colback=gray!20, colframe=gray!20, coltitle=black]
\textbf{P}$_{\text{Bas}}$: "Please analyze the input code snippet, and determine the most appropriate architectural tactic label from \{'audit', 'authenticate', 'heartbeat', 'pooling', 'scheduler', 'unrelated\}, providing your reasons. [Code snippet]"
\end{tcolorbox}

\item[$\bullet$ \textbf{P}$_{\text{ABas}}$](Augmented basic prompt): CoT \cite{DBLP:conf/nips/Wei0SBIXCLZ22} techniques can effectively guide LLMs through reasoning paths, reduce hallucinations, and improve detection accuracy. Here, we construct a three-step CoT to assist LLMs in performing ATs detection. First, we prompt the LLMs to identify the core behaviors and objectives of the input code snippet; second, we prompt the LLMs to detect whether the code snippet contains any ATs; and finally, we prompt the LLMs to provide the justification and explanation for the detection results.
\begin{tcolorbox}[colback=gray!20, colframe=gray!20, coltitle=black]
\textbf{P}$_{\text{ABas}}$: \newline
"\texttt{Step1: }Please understand the behavior and purpose of the input code snippet.\newline
\texttt{Step2: }Determine the unique true label of this code from \{'audit', 'authenticate', 'heartbeat', 'pooling', 'scheduler', 'unrelated\}.\newline
\texttt{Step3: }Please give your reasoning. [Code snippet]"
\end{tcolorbox}
\end{enumerate}

\subsection{Evaluation Metrics}
\subsubsection{Evaluating Accuracy}
The ATs detection task is inherently a classification task. To evaluate the accuracy of Prmt4TD, three standard evaluation metrics are introduced: \emph{Precision}, \emph{Recall}, and \emph{F1-Score}.  These metrics are widely used in the field of software engineering \cite{9825785}\cite{Sharifi2023} and have also been adopted in the baseline methods used in this study. Common parameters among the metrics include true positive (TP), true negative (TN), false positive (FP), and false negative (FN).

\textbf{\emph{Precision (Pre)}}: The proportion of samples predicted as positive that is actually positive. It indicates the classifier's accuracy in predicting each class. High \emph{Precision} means the classifier rarely misclassifies negative samples as positive, which is crucial in situations where false positives are costly.
\begin{equation}\footnotesize
Precision=\frac{TP}{TP + FP}\label{eq1}
\end{equation}

\textbf{\emph{Recall (Re)}}: The proportion of actual positive samples that are correctly predicted as positive by the classifier. It reflects the classifier's ability to identify all positive samples. High \emph{Recall} means the classifier can detect most positive samples, which is important when the cost of missing positive cases is high.
\begin{equation}\footnotesize
Recall=\frac{TP}{TP + FN}\label{eq2}
\end{equation}

\textbf{\emph{F1-Score (F1)}}: The \emph{F1-score} is the harmonic mean of precision and recall, providing a balance between the classifier's accuracy and completeness. A higher \emph{F1-score} indicates better performance, particularly suitable for datasets with imbalanced classes.
\begin{equation}
F1-score=\frac{2 \cdot \text{\emph{Precision}} \cdot \text{\emph{Recall}}}{\text{\emph{Precision}} + \text{\emph{Recall}}}\label{eq3}
\end{equation}

\subsubsection{Evaluating Comprehensibility}
\label{ex: comprehensibility}
In evaluating the comprehensibility of Prmt4TD for the ATs detection task, we opt for a human evaluation approach for the following reasons. First, ensuring the comprehensibility of ATs detection results is crucial for enhancing developers' trust in the outcomes. However, research in this area remains limited, with the lack of standardized benchmarks, such as datasets containing explanatory text that can be used as reference answers. Second, numerous studies have demonstrated that the outputs of LLMs exhibit significant diversity, rendering evaluation methods such as BLEU, which rely on string matching, less effective in assessing the performance of LLMs \cite{touvron2023llamaopenefficientfoundation}\cite{10.1145/3695993}. Drawing from methods used in evaluating the comprehensibility of LLMs-generated review comments in the code review domain \cite{10.1145/3695993}, we devise a three-category evaluation approach for human assessors to evaluate Prmt4TD's comprehensibility. This evaluation method is based on "clarity", including \emph{Clear}, \emph{Neutral}, or \emph{Unclear}, with the specific evaluation criteria shown in \Cref{tab: clarity}.

\begin{table}[htbp]
\scriptsize
\centering
\renewcommand\arraystretch{1.25}
\caption{\textbf{The different degrees of clarity for evaluating comprehensibility}}
\label{tab: clarity}
\begin{tabular}{p{1.5cm}p{6cm}}
\hline
\textbf{Clarity} & \textbf{Description} \\ \hline
\emph{Clear}            & The ATs detection comments generated by LLMs can accurately identify the architectural tactic present in the input code snippet (if any) and provide reasonable explanations. \\
\emph{Unclear}          & The ATs detection comments generated by LLMs fail to clearly identify the architectural tactic in the input code snippet or contain logical errors. This includes cases where inexistent architectural tactics are highlighted or where existing tactics are overlooked.  \\
\emph{Neutral}          & Other cases are categorized as neutral.\\ \hline
\end{tabular}
\vspace{-0.0ex}
\end{table}

\subsection{Small Models candidates}
\label{ex: ML models}
We select three ML models, SVM \cite{joachims1999transductive}, Adaboost \cite{freund1996experiments}, and Bagging \cite{breiman1996bagging} (cf. \Cref{ex: baselinesML}) and two DL models, BERT \cite{devlin2019bertpretrainingdeepbidirectional} and CodeBERT \cite{feng2020codebertpretrainedmodelprogramming}, as candidates for smaller models. Previous studies \cite{7270338}\cite{math8050851} have demonstrated that these ML models perform well in tasks such as ATs detection, which is a form of code classification. Additionally, these models represent a range of similar models within their respective categories: discriminative classifiers, boosting ensemble classifiers, and parallel ensemble classifiers.
Keim et al. fine-tune BERT to identify ATs in code and suggest that, for such tasks, the key prerequisite is ensuring that the chosen model is appropriate for the specific type of input data \cite{10.1007/978-3-030-58923-3_15}. Therefore, we also select BERT and CodeBERT as alternative options.

\begin{itemize}
    \item \textbf{BERT}: BERT is a DL model based on a bidirectional encoder, specifically designed for natural language processing tasks \cite{devlin2019bertpretrainingdeepbidirectional}.
    \item \textbf{CodeBERT}: CodeBERT is a DL model based on a self-attention mechanism, specifically designed for code understanding and analysis \cite{feng2020codebertpretrainedmodelprogramming}.
\end{itemize}

\subsection{Implementation Details}
We split the balanced dataset into training and test sets at an 8:2 ratio. The training set is used to train the small models, while the test set is employed to evaluate the effectiveness of Prmt4TD.  Considering generalization, except for the Tactic Det. \cite{7270338} and BERT4TD \cite{10.1007/978-3-030-58923-3_15} baseline methods, which are configured with the parameter settings reported in the papers, all other methods use their default parameter settings.

\section{Results and Analysis}
\label{sec:analyis}
\subsection{RQ1: Selecting Small Models}
\label{rq1}
We conducted experiments on the balanced dataset to investigate which type of small model would provide better support for Prmt4TD. The results are shown in \Cref{tab: Smodel}. Prmt4TD with CodeBERT achieves the best performance, excelling in the detection of four ATs. In comparison, Prmt4TD using AdaBoost, Bagging, and BERT outperforms in the detection of two ATs each. On the other hand, Prmt4TD with SVM demonstrates average performance, with no significant advantage in detecting any of the ATs. In terms of the comprehensive metric \emph{F1-score}, CodeBERT outperforms SVM, AdaBoost, Bagging, and BERT by an average of 5\%, 5\%, 9\%, and 4\%, respectively. This demonstrates that using CodeBERT as a plug-in model for Prmt4TD significantly enhances its generalization ability and adaptability when performing ATs detection task. Furthermore, the superior performance of CodeBERT compared to BERT validates Keim et al.'s hypothesis, which suggests that the alignment between language models and the type of input data can provide more effective support for downstream tasks \cite{10.1007/978-3-030-58923-3_15}.

\begin{table*}[htbp]
\scriptsize
\renewcommand\arraystretch{1.3}
\caption{\textbf{The Comparison of Various Small Models in The Balanced Dataset}}
\label{tab: Smodel}
\begin{tabular}{llllllllllllllll}
\hline
                 & \multicolumn{3}{c}{SVM}                  & \multicolumn{3}{c}{AdaBoost}             & \multicolumn{3}{c}{Bagging}              & \multicolumn{3}{c}{BERT}                 & \multicolumn{3}{c}{CodeBERT}                                       \\ \cmidrule(lr){2-4} \cmidrule(lr){5-7} \cmidrule(lr){8-10} \cmidrule(lr){11-13} \cmidrule(lr){14-16} 
                 & \textit{Pre} & \textit{Re} & \textit{F1} & \textit{Pre} & \textit{Re} & \textit{F1} & \textit{Pre} & \textit{Re} & \textit{F1} & \textit{Pre} & \textit{Re} & \textit{F1} & \textit{Pre} & \textit{Re} & \textit{F1}                           \\ \hline
Audit Trail      & 1.00         & 0.70        & 0.82        & 1.00         & 0.80        & \cellcolor[HTML]{C0C0C0}\textbf{0.89}        & 1.00         & 0.80        & \cellcolor[HTML]{C0C0C0}\textbf{0.89}        & 1.00         & 0.80        & \cellcolor[HTML]{C0C0C0}\textbf{0.89}        & 1.00         & 0.80        & \cellcolor[HTML]{C0C0C0}\textbf{0.89}                                  \\
Authentication   & 1.00         & 0.90        & 0.95        & 1.00         & 0.90        & 0.95        & 1.00         & 0.80        & 0.89        & 1.00         & 1.00        & \cellcolor[HTML]{C0C0C0}\textbf{1.00}        & 1.00         & 0.90        & 0.95                                  \\
Heartbeat        & 1.00         & 0.80        & 0.89        & 1.00         & 0.80        & 0.89        & 1.00         & 0.60        & 0.75        & 1.00         & 0.70        & 0.82        & 1.00         & 0.90        & \cellcolor[HTML]{C0C0C0}\textbf{0.95}                                  \\
Pooling & 1.00         & 0.90        & 0.95        & 1.00         & 1.00        & \cellcolor[HTML]{C0C0C0}\textbf{1.00}        & 1.00         & 1.00        & \cellcolor[HTML]{C0C0C0}\textbf{1.00}        & 1.00         & 1.00        & \cellcolor[HTML]{C0C0C0}\textbf{1.00}        & 1.00         & 1.00        & \cellcolor[HTML]{C0C0C0}\textbf{1.00}                                  \\
Scheduling       & 1.00         & 0.90        & 0.95        & 0.89         & 0.80        & 0.84        & 1.00         & 0.80        & 0.89        & 0.90         & 0.90        & 0.90        & 1.00         & 1.00        & \cellcolor[HTML]{C0C0C0}\textbf{1.00}                                  \\ \hline
\textit{avg}     & 1.00         & 0.84        & 0.91        & 0.98         & 0.86        & 0.91        & 0.98         & 0.80        & 0.87        & 0.98         & 0.88        & 0.92        & 1.00         & 0.92        & \cellcolor[HTML]{C0C0C0}\textbf{0.96} \\ \hline
\end{tabular}
\end{table*}

\begin{tcolorbox}[width=\linewidth, boxrule=1.0pt, left=2pt, right=2pt, top=2pt, bottom=2pt, colback=gray!0]
 \textbf{Summary for RQ1:}
Among the different small models, CodeBERT provides superior performance for Prmt4TD, maintaining strong accuracy across different types of ATs detection tasks. Therefore, we choose to use CodeBERT as the plug-in model for Prmt4TD, supporting the construction of in-context prompts.
\end{tcolorbox}

\subsection{RQ2: Evaluating Accuracy of Prmt4TD}\label{rq2}
We employed DeepSeek-v3 \footnote{\url{https://github.com/deepseek-ai/DeepSeek-V3}}, an advanced model known for its outstanding performance and the best value in the market \cite{deepseekai2024deepseekv3technicalreport}, as the LLM for ATs detection in prmt4TD. During the experimentation, we replicated baseline methods using both the balanced dataset and the Hadoop dataset. This is necessary because most of the baseline methods in the original papers are evaluated on datasets that differed from the ones used in our study, and these datasets are not accessible. Additionally, due to the broken link to the open-source implementation of the Tactic Det. \cite{7270338}  method, we replicated this method based on the details provided in the paper.

\Cref{tab: results of diff method on dataset} presents a comparison of the detection accuracy between Prmt4TD and baseline methods. Across all evaluation metrics, Prmt4TD demonstrates superior performance in detecting ATs, showcasing its high accuracy and comprehensiveness in this task. For example, in terms of \emph{precision} and \emph{recall}, Prmt4TD outperforms the next-best method, Bagging, by an average of 9\% and 16\%, respectively. Besides, in the overall \emph{F1-score}, Prmt4TD exceeds other baseline methods by a margin ranging from a minimum of 13\% to a maximum of 23\%. To evaluate the performance of Prmt4TD in detecting ATs in large-scale industrial projects, we conducted experiments using the Hadoop case. The results are presented in \Cref{tab: results of diff method in Hadoop}.  Prmt4TD demonstrates superior performance in detecting three different ATs, while SVM, AdaBoost, and Tactic. det outperform in detecting one AT, respectively. In terms of the overall \emph{F1-score}, Prmt4TD consistently outperforms the other baseline methods by at least 4\% and up to 23\%. This further highlights the outstanding performance and generalization capability of Prmt4TD in performing ATs detection tasks.

Additionally, we observe that the performance of all classifiers significantly decreased in the Hadoop case. This could be due to the imbalanced sample distribution and relatively ambiguous sample boundaries in the dataset, making the detection of ATs in Hadoop code more challenging. For instance, in the detection of the Audit Trail tactic, which has only four labeled data samples, Prmt4TD shows a 41\% decrease in \emph{F1-score} compared to the balanced dataset. Even the relatively better-performing SVM and Tactic Det. baseline methods experienced \emph{F1-score} decreases of 17\% and 35\%, respectively. Furthermore, we notice that the worst-performing SVM baseline method completely fails to detect the heartbeat tactic. This phenomenon is not only related to the aforementioned factors but also further confirms the limitations of ML models of the SVM type: their high sensitivity to data quality and relatively limited generalization capability. In contrast, Prmt4TD still maintains the best performance, demonstrating that Prmt4TD is more robust and adaptable compared to other baseline methods when facing changes in data distribution.
\begin{table*}[htbp]
\scriptsize
\renewcommand\arraystretch{1.3}
\caption{\textbf{Results of Different Classifiers for Detecting Architectural Tactics in The Balanced Dataset}}
\label{tab: results of diff method on dataset}
\begin{tabular}{lccccccccccccccccccc}  
\toprule
                 & \multicolumn{3}{c}{SVM} & \multicolumn{3}{c}{AdaBoost} & \multicolumn{3}{c}{Bagging}  & \multicolumn{3}{c}{BERT4TD} & \multicolumn{3}{c}{Tactic Det.} & \multicolumn{3}{c}{Prmt4TD}  \\  
\cmidrule(lr){2-4} \cmidrule(lr){5-7} \cmidrule(lr){8-10} \cmidrule(lr){11-13} \cmidrule(lr){14-16} \cmidrule(lr){17-19}  
                 & \textit{Pre} & \textit{Re} & \textit{F1} & \textit{Pre} & \textit{Re} & \textit{F1}                           & \textit{Pre} & \textit{Re} & \textit{F1}                           & \textit{Pre} & \textit{Re} & \textit{F1} & \textit{Pre} & \textit{Re} & \textit{F1}                           & \textit{Pre} & \textit{Re} & \textit{F1}    \\  
\midrule
Audit Trail      & 1.00         & 0.40        & 0.57        & 0.88         & 0.70        & 0.78        & 0.88         & 0.70        & 0.78        & 0.89         & 0.80        & 0.84        & 1.00         & 0.60         & 0.75        & 1.00         & 0.80        & \cellcolor[HTML]{C0C0C0}\textbf{0.89} \\
Authentication   & 1.00         & 0.80        & 0.89        & 0.73         & 0.80        & 0.76        & 0.80         & 0.80        & 0.80        & 0.75         & 0.90        & 0.82        & 1.00         & 0.60        & 0.75        & 1.00         & 0.90        & \cellcolor[HTML]{C0C0C0}\textbf{0.95} \\
Heartbeat        & 1.00         & 0.60        & 0.75        & 0.88         & 0.70        & 0.78        & 0.88         & 0.70        & 0.78        & 0.78         & 0.70        & 0.74        & 0.75         & 0.30        & 0.43        & 1.00         & 0.90        & \cellcolor[HTML]{C0C0C0}\textbf{0.95} \\
Pooling & 1.00         & 0.70        & 0.82        & 0.91         & 1.00        & 0.95        & 1.00         & 0.90        & 0.95        & 0.82         & 0.90        & 0.86        & 0.86         & 0.60        & 0.71        & 1.00         & 1.00        & \cellcolor[HTML]{C0C0C0}\textbf{1.00} \\
Scheduling       & 1.00         & 0.80        & 0.89        & 0.88         & 0.70        & 0.78        & 1.00         & 0.70        & 0.82        & 0.75         & 0.90        & 0.82        & 1.00         & 1.00        & \cellcolor[HTML]{C0C0C0}\textbf{1.00}        & 1.00         & 1.00        & \cellcolor[HTML]{C0C0C0}\textbf{1.00} \\ \hline
\textit{avg}     & 1.00         & 0.66        & 0.78        & 0.86         & 0.78        & 0.81        & 0.91         & 0.76        & 0.83        & 0.80         & 0.84        & 0.82        & 0.92         & 0.62        & 0.73        & 1.00         & 0.92        & \cellcolor[HTML]{C0C0C0}\textbf{0.96} \\ \hline
\end{tabular}
\end{table*}

\begin{table*}[htbp]
\scriptsize
\renewcommand\arraystretch{1.3}
\caption{\textbf{Results of Different Classifiers for Detecting Architectural Tactics in Hadoop}}
\label{tab: results of diff method in Hadoop}
\begin{tabular}{lllllllllllllllllll}
\hline
                 & \multicolumn{3}{c}{SVM}                  & \multicolumn{3}{c}{AdaBoost}                                       & \multicolumn{3}{c}{Bagging}                                        & \multicolumn{3}{c}{BERT4TD}              & \multicolumn{3}{c}{Tactic Det.}                                    & \multicolumn{3}{c}{Prmt4TD}                                        \\ \cmidrule(lr){2-4} \cmidrule(lr){5-7} \cmidrule(lr){8-10} \cmidrule(lr){11-13} \cmidrule(lr){14-16} \cmidrule(lr){17-19} 
                 & \textit{Pre} & \textit{Re} & \textit{F1} & \textit{Pre} & \textit{Re} & \textit{F1}                           & \textit{Pre} & \textit{Re} & \textit{F1}                           & \textit{Pre} & \textit{Re} & \textit{F1} & \textit{Pre} & \textit{Re} & \textit{F1}                           & \textit{Pre} & \textit{Re} & \textit{F1}                           \\ \midrule
Audit Trail      & 1.00         & 0.25        & \cellcolor[HTML]{C0C0C0}\textbf{0.40}        & 0.17         & 0.25        & 0.20                                  & 0.50         & 0.25        & 0.33                                  & 0.5          & 0.25        & 0.33        & 1.00         & 0.25        & \cellcolor[HTML]{C0C0C0}\textbf{0.40} & 0.20         & 0.25        & 0.22                                  \\
Authentication   & 0.82         & 0.36        & 0.50        & 0.73         & 0.44        & 0.55                                  & 0.92         & 0.48        & 0.63                                  & 0.91         & 0.40        & 0.56        & 1.00         & 0.12        & 0.21                                  & 0.94         & 0.64        & \cellcolor[HTML]{C0C0C0}\textbf{0.76} \\
Heartbeat        & 0.00         & 0.00        & 0.00        & 0.45         & 0.54        & 0.49                                  & 0.70         & 0.58        & \cellcolor[HTML]{C0C0C0}\textbf{0.64} & 0.76         & 0.54        & 0.63        & 1.00         & 0.08        & 0.15                                  & 0.68         & 0.54        & 0.60                                  \\
Pooling & 1.00         & 0.02        & 0.05        & 0.98         & 0.51        & \cellcolor[HTML]{C0C0C0}\textbf{0.67} & 1.00         & 0.02        & 0.05                                  & 1.00         & 0.21        & 0.35        & 0.95         & 0.24        & 0.38                                  & 1.00         & 0.28        & 0.44                                  \\
Scheduling       & 0.69         & 0.59        & 0.63        & 0.77         & 0.49        & 0.59                                  & 0.85         & 0.51        & 0.64                                  & 0.82         & 0.59        & 0.68        & 0.75         & 0.31        & 0.44                                  & 0.83         & 0.63        & \cellcolor[HTML]{C0C0C0}\textbf{0.72} \\ \hline
\textit{avg}     & 0.70         & 0.24        & 0.32        & 0.62         & 0.45        & 0.50                                  & 0.79         & 0.37        & 0.46                                  & 0.80         & 0.40        & 0.51        & 0.94         & 0.20        & 0.32                                  & 0.73         & 0.47        & \multicolumn{1}{c}{\cellcolor[HTML]{C0C0C0}\textbf{0.55}}              \\ \hline
\end{tabular}
\end{table*}

\begin{tcolorbox}[width=\linewidth, boxrule=1.0pt,
  left=2pt, right=2pt, top=2pt, bottom=2pt, colback=gray!0]
 \textbf{Summary for RQ2:} 
The overall performance of prmt4TD outperforms other baseline methods, as evidenced by its higher \emph{F1-scores} across different datasets.
\end{tcolorbox}

\subsection{RQ3: Evaluating Comprehensibility of Prmt4TD}\label{rq3}
\begin{figure*}[htbp]
    \begin{center}
    \includegraphics[width=1.0\textwidth]{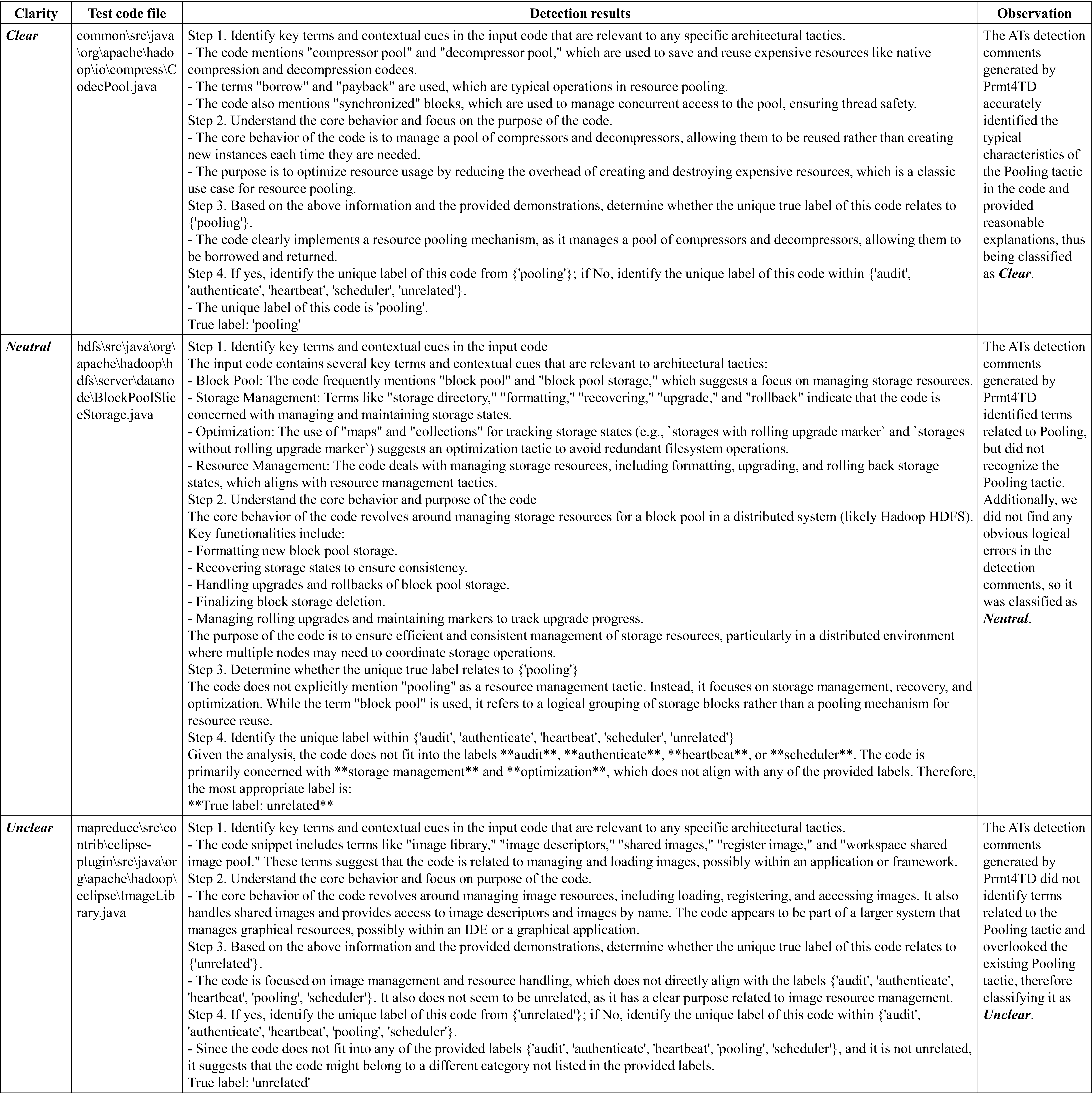}
    \vspace{-2.0ex}
    \end{center}
    \caption{Examples of Clarity in the Comprehensibility evaluation}
    \vspace{-4.0ex}
\label{fig: exampleClarity}
\end{figure*}
When the accuracy of ATs detection methods is suboptimal, evaluating the comprehensibility of the generated detection comments becomes impractical, as the detected results may inherently be erroneous. In this study, we compared Prmt4TD with baseline prompting frameworks (cf. \Cref{sec:baseline for Comprehensibility}). To mitigate any potential bias that the training dataset might introduce to Prmt4TD, we chose to evaluate the corresponding methods using all the positive samples from the Hadoop case. The comprehensibility evaluation task was independently performed by two evaluators (two developers). To ensure the objectivity and reliability of the evaluation, we concealed the specific ATs detection methods used to generate the detection comments and insights when presenting them to the evaluators. 
 \begin{table}[htbp]
\scriptsize
\centering
\caption{\textbf{Performance Evaluation of Various Prompting Framework in Hadoop}}
\label{tab: compare promts}
\renewcommand\arraystretch{1.3}
\begin{tabular}{p{1.3cm} p{0.23cm}p{0.28cm}p{0.26cm} p{0.23cm}p{0.28cm}p{0.26cm} p{0.23cm}p{0.28cm}p{0.26cm}}
\hline
                 & \multicolumn{3}{c}{\textbf{P}$_{\text{Bas}}$}                 & \multicolumn{3}{c}{\textbf{P}$_{\text{ABas}}$}                & \multicolumn{3}{c}{Prmt4TD}              \\ \cmidrule(lr){2-4} \cmidrule(lr){5-7} \cmidrule(lr){8-10} 
                 & \textit{Pre} & \textit{Re} & \textit{F1} & \textit{Pre} & \textit{Re} & \textit{F1} & \textit{Pre} & \textit{Re} & \textit{F1} \\ \hline
Audit Trail      & 0.14         & 0.50        & 0.22                                  & 0.08         & 0.50        & 0.14                                  & 0.20         & 0.50        & \cellcolor[HTML]{C0C0C0}\textbf{0.29} \\
Authentication   & 0.61         & 0.76        & 0.68                                  & 0.70         & 0.84        & \cellcolor[HTML]{C0C0C0}\textbf{0.76} & 0.70         & 0.76        & 0.73                                  \\
Heartbeat        & 0.69         & 0.75        & 0.72                                  & 0.82         & 0.58        & 0.68                                  & 0.82         & 0.75        & \cellcolor[HTML]{C0C0C0}\textbf{0.78} \\
Pooling & 0.93         & 0.46        & \cellcolor[HTML]{C0C0C0}\textbf{0.61} & 0.76         & 0.46        & 0.57                                  & 0.93         & 0.29        & 0.45                                  \\
Scheduling       & 0.68         & 0.75        & 0.71                                  & 0.70         & 0.74        & 0.72                                  & 0.76         & 0.81        & \cellcolor[HTML]{C0C0C0}\textbf{0.79} \\ \hline
\textit{avg}     & 0.61         & 0.64        & 0.59                                  & 0.61         & 0.62        & 0.57                                  & 0.68         & 0.62        & \cellcolor[HTML]{C0C0C0}\textbf{0.61} \\ \hline
\end{tabular}
\end{table}

\begin{figure}[htbp]
\begin{center}
    \includegraphics[width=1.0\linewidth]{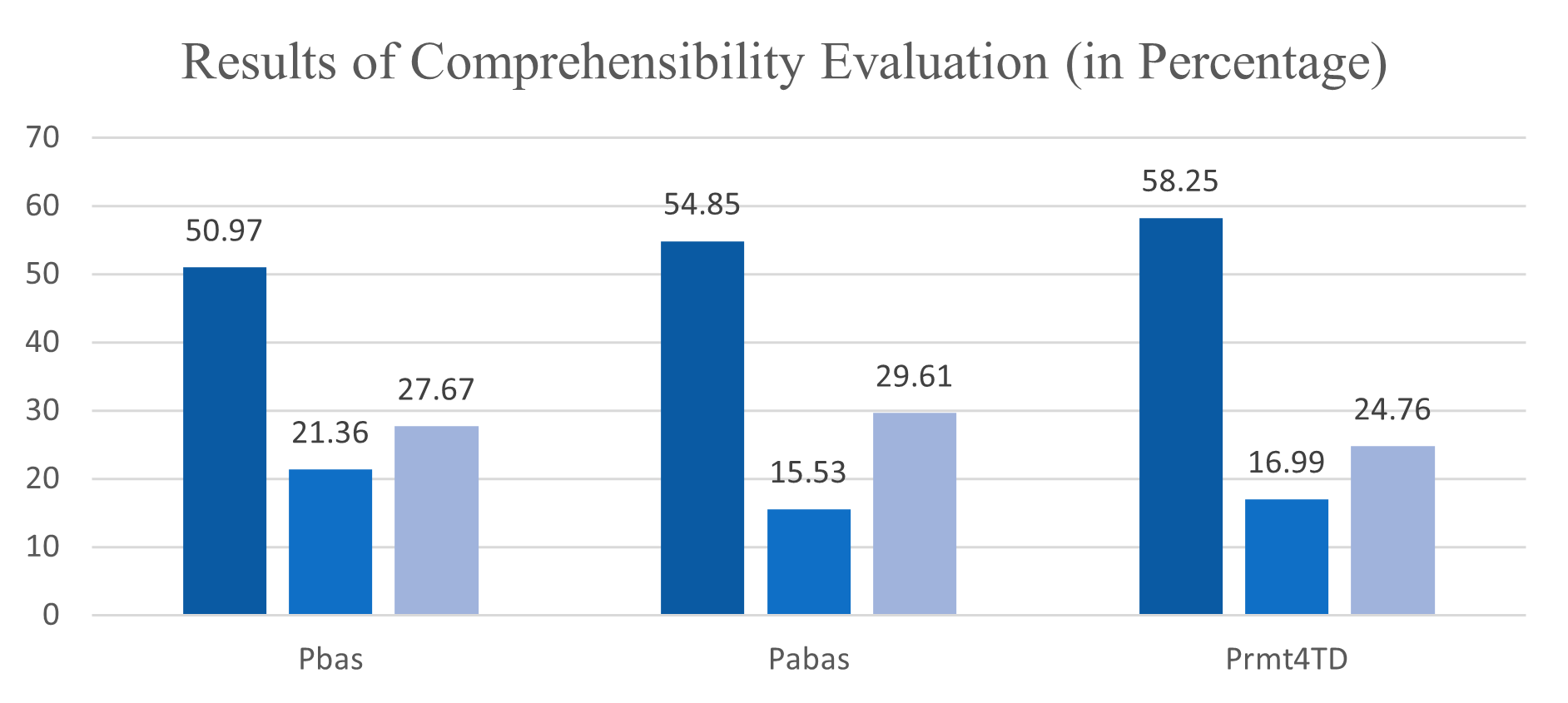}
    \vspace{-2.0ex}
\end{center}
\vspace{-4.0ex}
\caption{Performance in Terms of Comprehensibility of Detection Results (\colorbox{myblue1}{} \emph{Clear} 
\colorbox{myblue2}{} \emph{Neutral} 
\colorbox{myblue3}{} \emph{Unclear})}
\vspace{-2.0ex}
\label{fig: clarity}
\end{figure}

Specifically, we first conducted a performance evaluation of the detection methods, with the results presented in \Cref{tab: compare promts}. In terms of the \emph{F1-score}, Prmt4TD performs exceptionally well in detecting three ATs, while \textbf{P}$_{\text{Bas}}$ and \textbf{P}$_{\text{ABas}}$ excel in detecting one AT each. In the evaluation of the \emph{average F1- score}, Prmt4TD achieves the highest score of 61\%. For the comprehensibility evaluation, the first evaluator assessed the first 60\% of the test data, while the second evaluator assessed the last 60\%. The overlapping 10\% of the data was used for a chi-square test \cite{Pearson01071900} to assess the consistency between the evaluations of the two assessors. \Cref{fig: exampleClarity} presents visual examples of different degrees of clarity. The results of the chi-square test indicate that the P-value between the evaluations provided by the two assessors is greater than 0.05, confirming that there is no significant difference between their evaluation outcomes. \Cref{fig: clarity} illustrates the performance of prmt4TD compared to the baseline prompting frameworks in terms of the comprehensibility of the generated detection comments. Prmt4TD provides a higher number of "clear" detection comments and fewer "unclear" comments, outperforming all baseline methods.  By combining the generative and reasoning capabilities of LLMs with the COT framework we developed, Prmt4TD significantly enhances the comprehensibility of ATs detection comments.

\begin{tcolorbox}[width=\linewidth, boxrule=1.0pt, left=2pt, right=2pt, top=2pt, bottom=2pt, colback=gray!0]
\textbf{Summary for RQ3:} Compared to other baseline prompting frameworks, Prmt4TD excels in both the \emph{F1-score} and clarity metrics, effectively balancing accuracy and comprehensibility.
\end{tcolorbox}

\subsection{RQ4: Ablation Study}\label{rq3}
\begin{table*}[htbp]
\scriptsize
\centering
\caption{\textbf{Reuslts of the Ablation Study}}
\renewcommand\arraystretch{1.3}
\label{tab: ablation study}
\begin{tabular}{llllllllllllllll}
\hline
                 & \multicolumn{3}{c}{w/o Small Model}                                & \multicolumn{3}{c}{w/o CoT}                                        & \multicolumn{3}{c}{w/o Demonstration}                              & \multicolumn{3}{c}{\begin{tabular}[c]{@{}c@{}}Random \\ Demonstration\end{tabular}} & \multicolumn{3}{c}{Prmt4TD}                                        \\ \cmidrule(lr){2-4} \cmidrule(lr){5-7} \cmidrule(lr){8-10} \cmidrule(lr){11-13} \cmidrule(lr){14-16} 
                 & \textit{Pre} & \textit{Re} & \textit{F1}                           & \textit{Pre} & \textit{Re} & \textit{F1}                           & \textit{Pre} & \textit{Re} & \textit{F1}                           & \textit{Pre}                & \textit{Re}               & \textit{F1}               & \textit{Pre} & \textit{Re} & \textit{F1}                           \\ \midrule
Audit Trail      & 0.40         & 0.50        & \cellcolor[HTML]{C0C0C0}\textbf{0.44} & 0.25         & 0.25        & 0.25                                  & 0.29         & 0.50        & 0.36                                  & 0.20                        & 0.25                      & 0.22                      & 0.20         & 0.50        & 0.29                                  \\
Authentication   & 0.84         & 0.64        & 0.73                                  & 0.82         & 0.72        & 0.77                                  & 1.00         & 0.72        & \cellcolor[HTML]{C0C0C0}\textbf{0.84} & 0.94                        & 0.64                      & 0.76                      & 0.70         & 0.76        & 0.73                                  \\
Heartbeat        & 0.74         & 0.58        & 0.65                                  & 0.70         & 0.58        & 0.64                                  & 0.68         & 0.62        & 0.65                                  & 0.72                        & 0.54                      & 0.62                      & 0.82         & 0.75        & \cellcolor[HTML]{C0C0C0}\textbf{0.78} \\
Pooling & 1.00         & 0.22        & 0.37                                  & 1.00         & 0.29        & \cellcolor[HTML]{C0C0C0}\textbf{0.45} & 0.92         & 0.28        & 0.43                                  & 0.96                        & 0.27                      & 0.42                      & 0.93         & 0.29        & \cellcolor[HTML]{C0C0C0}\textbf{0.45} \\
Scheduling       & 0.79         & 0.65        & 0.71                                  & 0.80         & 0.66        & 0.73                                  & 0.84         & 0.60        & 0.70                                  & 0.81                        & 0.63                      & 0.71                      & 0.76         & 0.81        & \cellcolor[HTML]{C0C0C0}\textbf{0.79} \\ \hline
\textit{avg}     & 0.75         & 0.52        & 0.58                                  & 0.71         & 0.50        & 0.57                                  & 0.75         & 0.54        & 0.60                                  & 0.73                        & 0.47                      & 0.55                      & 0.68         & 0.62        & \cellcolor[HTML]{C0C0C0}\textbf{0.61} \\ \hline
\end{tabular}
\end{table*}

To optimize costs while maintaining a representative subset of data, we continue to use the Hadoop dataset, which was originally employed to evaluate the comprehensibility of Prmt4TD. During the experimentation process, in the interest of fairness, the test data used in each ablation experiment remains constant. The experimental results are presented in \Cref{tab: ablation study}. We first attempted to evaluate the impact of the small model on the performance of Prmt4TD. Evidently, without the small model integrated as a plug-in, the \emph{F1-score} experiences a significant decline, averaging a 3\% reduction compared to Prmt4TD. This observation highlights the importance of integrating the small model into Prmt4TD. By incorporating the knowledge provided by the small model, it is possible to transfer the task-specific knowledge learned by the small model to DeepSeek-v3 in a manner similar to knowledge distillation. After removing the small model, DeepSeek-v3 is unable to capture the uncertainty of CodeBERT in the task, making it difficult to determine when to overwrite previous predictions. We then attempted to remove CoT from Prmt4TD, which had a negative impact on its performance.  The absence of a reasoning path may hinder DeepSeek-v3's ability to accurately capture the logical relationships within the task, leading to a limited understanding of the task and, consequently, a decrease in accuracy. When we completely removed the demonstrations from ICL, DeepSeek-v3 relied on the knowledge of the small model and CoT for inference analysis on the test inputs. In this case, the model's performance in terms of \emph{F1-score} still falls short compared to Prmt4TD. We believe this outcome results from the demonstrations in ICL helping DeepSeek-v3 learn and calibrate the confidence of the predictions made by the small model, thereby improving the accuracy of the final prediction. Additionally, we also experimented with providing random demonstrations for ICL, which resulted in a significant performance decline. Compared to Prmt4TD, the average performance drops by 6\%. These demonstrations, which may be unrelated to the test inputs, hinder the sufficiency and effectiveness of the information in ICL. This inadequacy interferes with DeepSeek-V3's ability to learn and comprehend the tasks, ultimately resulting in the observed decline in ICL performance.

\begin{tcolorbox}[width=\linewidth, boxrule=1.0pt, left=2pt, right=2pt, top=2pt, bottom=2pt, colback=gray!0]
\textbf{Summary for RQ4:} The results of the ablation experiments demonstrate that integrating the small model, CoT, and carefully selected demonstrations significantly enhances the performance of Prmt4TD in detecting ATs in code.
\end{tcolorbox}

\section{Discussion}
\label{sec:discussion}
\subsection{Automated Detection of ATs in Code Using LLMs}
Traditional ML-based approaches for automated detection of ATs in code often focus on improving the accuracy of detection results while overlooking the comprehensibility of those results. This lack of comprehensibility can introduce additional workload for developers when confirming the detected results. With the advent of the AI era, LLMs have demonstrated exceptional performance, powerful reasoning, and generation capabilities. This paper introduces LLMs to the task of automated detection of ATs in code, leveraging prompt engineering augmented by small models. Experimental results confirm the significant potential of LLMs in performing such tasks, providing effective support for automated ATs detection. With the continuous advances in research on LLMs and their applications, we anticipate that these developments will provide new opportunities for the effectiveness and advancements of the proposed method. The proposed method involves training a small model to learn task-specific knowledge, which is then transferred to LLMs through ICL. ICL enables to triggering implicit fine-tuning of the LLMs, providing precise domain knowledge for ATs detection tasks while reducing costs. Additionally, the CoT technique helps the LLM establish clear reasoning paths, thereby enhancing its performance in task execution. Additionally, in transformers, the attention weights are computed by multiplying the input size by the output size. The application of CP techniques aids in selecting demonstrations to reduce the number of demonstration categories in ICL, which increases the probability of the model selecting the correct category. This, in turn, improves prediction performance and reduces energy consumption \cite{randl2024cicle}. The integration of these components suggests that the proposed method holds significant potential for enhancing the performance of automated ATs detection in code. Compared to other methods, Prmt4TD offers easy scalability to accommodate additional ATs with lower cost and resource requirements. It effectively transfers the computational and time costs of using LLMs to smaller models while providing comprehensibility of the results. This makes it highly effective in detecting ATs in code, demonstrating good performance in this task. As architectural design decisions that reflect system quality attributes, ATs play a critical role throughout the architecture life cycle, making the detection of ATs in code an essential endeavor. If accuracy and comprehensibility cannot be ensured, the inefficiencies inherent in manually reviewing ATs in code will significantly undermine the value brought by software architecture empowered by LLMs.

\subsection{Data for Automated Detection of ATs in Code}
In the context of automated detection of ATs in code, data plays a crucial role in the training of models. The nature of the data directly determines the extent to which a model can identify and distinguish ATs accurately. Firstly, the dataset should be capable of effectively representing the features of various ATs and include a diversity of code implementations for these ATs. A lack of diversity in the dataset could lead to model over fitting, thereby reducing its generalization ability. In this study, a balanced dataset is used during model training, where each AT is represented by 50 implementation examples and 50 unrelated code snippets. Although this dataset size may not fully capture all the features of ATs, it provides an initial foundation for the automated detection of ATs in code. Furthermore, the accuracy of data labeling significantly impacts the detection performance of models. In validating Prmt4TD, we employ a Hadoop case study. Taking the labeling of the Audit Trail architectural tactic as an example, there were four data points in the case labeled as Audit Trail. The experimental results indicate that all classifiers perform poorly in detecting this AT. The Tactic Det. \cite{7270338}, proposed the Hadoop dataset, showes the best performance, but the \emph{F1-score} is still only 40\%. Upon reviewing these four data points, we find that some of them contained multiple ATs with unclear boundaries for identification. This could be one of the reasons behind the poor detection performance of the classifiers. Expanding and improving the quality of the ATs dataset is one of our future research directions. Building a comprehensive, accurate, and continuously updated dataset of ATs is a key step in advancing the research on automated detection of ATs in code. For multi-class classification tasks with the coexistence of multiple labels, accurately labeling the data will be an area for further exploration. Additionally, the preprocessing methods applied to the data are a critical factor influencing the performance of model detection, especially in the case of LLMs. If the preprocessing method is inadequate, such as when retained data features are inaccurate or missing, it may lead to false negatives and false positives, thus affecting the effectiveness of detection. On the other hand, retaining excessive redundant data or overly long data may interfere with the model's ability to learn relevant patterns, thereby reducing its performance. Therefore, we believe that exploring methods to preprocess data in a way that accurately captures the essential features while minimizing the data volume, thereby improving the performance of LLMs in downstream tasks, presents an interesting and promising research direction.

\section{Threats to Validity}
\label{sec:ttv}
In this section, we analyze potential threats to the validity of this study.

\textbf{Internal Validity.}
LLMs are typically trained on vast amounts of publicly available data from the web, and it is possible that the dataset used in this study is included in such data. However, the primary objective of this paper is to enhance the performance of LLMs in ATs detection tasks through the use of a small model-augmented prompting framework (Prmt4TD). To assess the effectiveness of the proposed framework, we conducted ablation experiments, which demonstrated that Prmt4TD successfully achieves this goal.

\textbf{External Validity.} 
Different versions of the same series of LLMs may exhibit varying performance on the same ATs detection task due to differences in their parameter settings. When reproducing the methods outlined in this paper, the choice of LLM version may lead to variations in the results. Therefore, we recommend using the same version of the LLM as in this paper to ensure the comparability and consistency of the experimental outcomes.

\textbf{Construct Validity.}
Prompts play a crucial role in guiding LLMs to perform downstream tasks effectively. Therefore, this paper introduces a small model-augmented prompting framework, which serves as a prompt template to guide LLMs in executing ATs detection tasks, embedding essential contextual information. Experimental results demonstrate that the proposed prompting framework can effectively perform ATs detection, ensuring both the accuracy of the detection and the comprehensibility of the results.

\textbf{Conclusion Validity.}
The dataset used in this study consists entirely of Java code, and thus the effectiveness of Prmt4TD in other programming languages has not been validated. Although the proposed prompting framework is theoretically applicable to multilingual scenarios, its practical performance in different programming languages still requires further validation. We will continue to explore this aspect in future research, examining its applicability and performance across other programming languages.

\section{Conclusion}
\label{sec:conclusion}
This paper presents a customized prompting framework, Prmt4TD, designed specifically for ATs detection tasks. The proposed method achieves both accuracy and comprehensibility, offering developers a convenient way to confirm detection results. Prmt4TD has shown outstanding performance across different datasets, outperforming other baseline methods, which highlights its stability and generalizability. In the future, we plan to continue the evaluation and refinement of Prmt4TD in practical scenarios, with the aim of providing robust support for architects and developers.

\section*{Data Availability}
The data related to Prmt4TD in this study are available at https://github.com/llc202jy/Prmt4TD.

\section*{Acknowledgments}
This work is supported by the National Natural Science Foundation of China (No.62302210, No.62202219), the Jiangsu Provincial Key Research and Development Program (No.BE202-1002-2), the Natural Science Foundation of Jiangsu Province (No.BK20241195), and the Innovation Project and Overseas Open Project of State Key Laboratory for Novel Software Technology (Nanjing University) (ZZKT2024A18, ZZKT2024-B07, KFKT2023A09, KFKT2023A10, KFKT2024A02, KFKT-2024A13, KFKT2024A14).

\bibliographystyle{cas-model2-names}
\bibliography{references}

\end{document}